\documentclass[%
 aip,
 amsmath,amssymb,
 reprint,%
]{revtex4-1}
\usepackage{subcaption}
\usepackage{physics}
\usepackage{graphicx}% Include figure files
\usepackage{dcolumn}% Align table columns on decimal point
\usepackage{bm}% bold math
\usepackage[utf8]{inputenc}
\usepackage[T1]{fontenc}
\usepackage{mathptmx}
\usepackage{etoolbox}
\usepackage{color,graphicx,slashed}
\usepackage[table]{xcolor}
\usepackage{url}
\usepackage{ulem}
\usepackage{xcolor}
\usepackage{comment}
\usepackage{hyperref}
\usepackage{amssymb}

\newcommand{\pina}[1]{{\color{magenta} #1}}

\newcommand{\lcpq}{Univ Toulouse, CNRS, Laboratoire de Chimie et Physique Quantiques, Toulouse, France}
\newcommand{\lpt}{Univ Toulouse, CNRS, Laboratoire de Physique Théorique, Toulouse, France}
\newcommand{\etsf}{ European Theoretical Spectroscopy Facility, Toulouse }
\newcommand{\TD}[1]{\color{olive} #1 \color{black}}

\makeatletter
\def\@email#1#2{%
 \endgroup
 \patchcmd{\titleblock@produce}
  {\frontmatter@RRAPformat}
  {\frontmatter@RRAPformat{\produce@RRAP{*#1\href{mailto:#2}{#2}}}\frontmatter@RRAPformat}
  {}{}
}%
\makeatother
\begin{document}

\preprint{AIP/123-QED}

\title[Algebraic Diagrammatic Construction of the Multichannel Dyson Equation]{Algebraic Diagrammatic Construction of the Multichannel Dyson Equation}
% Force line breaks with \\
\author{T. Demartini}
 \affiliation{CEA,DAM,DIF, 91297 Arpajon, France}
\affiliation{Université Paris-Saclay, CEA, Laboratoire Matière en Conditions Extrêmes, 91680 Bruyères-Le-Châtel, France}
\author{J.A. Berger}
\affiliation{\lcpq}%
\affiliation{\etsf}
\author{G. Blanchon}%
\affiliation{CEA,DAM,DIF, 91297 Arpajon, France}
\affiliation{Université Paris-Saclay, CEA, Laboratoire Matière en Conditions Extrêmes, 91680 Bruyères-Le-Châtel, France}
\author{T. Duguet}
\affiliation{Université Paris-Saclay, CEA, IRFU, 91191 Gif-sur-Yvette, France}%
\author{D. Lacroix}
\affiliation{Université Paris-Saclay, CNRS/IN2P3, IJCLab, 91405 Orsay, France}%
\author{P. Romaniello}
\affiliation{\lpt}%
\affiliation{\etsf}

\author{V. Som\`a}
\affiliation{Université Paris-Saclay, CEA, IRFU, 91191 Gif-sur-Yvette, France}%

\date{\today}% It is always \today, today,
             %  but any date may be explicitly specified

\begin{abstract}

The multichannel Dyson equation (MCDE) was recently introduced as a new approximation scheme to compute the one-body Green’s function [Riva \textit{et al.}, \textit{Phys. Rev. Lett.} \textbf{131}, 216401 (2023)] in many-body systems. The physical content of this novel approximation scheme is further clarified by recovering it from (an extended version of) the algebraic diagrammatic construction (ADC) truncation scheme. It is thus demonstrated that the MCDE approximation lies in between the so-called ADC(2) and ADC(3) truncations of the dynamical self energy. Building on this clarification, the MCDE approximation is tested on the periodic one-dimensional Hubbard model with 4-, 6- and 8-site lattices and shown to deliver an improved treatment over ADC(2) of both the quasiparticle peaks and the so-called satellites in the spectral strength distribution.
\end{abstract}

\maketitle

\section{Introduction}
\label{sec:into}

Charged excitations play a central role in the response of matter to external probes and are a key concept across quantum chemistry, nuclear physics, and condensed matter physics, from redox potentials and spectroscopic signatures in molecules, to collective excitations in atomic nuclei, to band structure and correlation-driven phenomena in solids. Achieving accurate theoretical descriptions of these processes requires capturing the same underlying many-body physics that these communities describe in different languages -- configuration mixing, screening, and strong correlation. A particularly demanding aspect is the description of the incoherent part of the excitation spectrum -- the satellites -- which arises beyond the quasi-particle picture and remains a challenge for most standard approaches. 
In this context, theoretical frameworks based on so-called Green’s functions are particularly valuable, as they provide a direct connection to spectroscopic observables. The one-body Green's function (1-GF), for instance, encodes information about charged excitations. The present work addresses the challenges described above by employing a recently introduced approximation scheme for calculating the one-body Green's function, in which quasiparticles and satellites are treated on the same footing: the multichannel Dyson equation (MCDE) scheme, which explicitly couples various n-body Green’s functions through a multichannel static self-energy~\cite{2022ScPP...12...93R,PhysRevLett.131.216401,PhysRevB.110.115140,PhysRevB.111.195133,10.1063/5.0291280}. The use of a static self-energy is a key feature, as it guarantees the absence of unphysical solutions~\cite{Lani_2012,Berger_2014,Stan_2015}. From a general perspective, a hierarchy of (n,s)-MCDE approximations can be characterized by two integers, i.e. (i) the change in particle number ($s$) between initial and final states and (ii) the highest-order ($n$) Green’s function involved. Finally, any (n,s)-MCDE can be recast as an eigenvalue problem involving an effective Hamiltonian, solvable with standard numerical techniques~\cite{RHaydock_1972, PhysRevB.67.085307, 10.1145/1089014.1089019}. In this work we focus on the (3,1)-MCDE, which allows for the calculations of charged excitations and related spectra. To date, the (3,1)-MCDE truncation scheme  coupling the 1-body Green's function to the two-electron-one-hole ($2e1h$) and one-electron-two-hole ($1e2h$) channels of the 3-body Green's function has been applied to the Hubbard dimer~\cite{PhysRevLett.131.216401}, the extended Hubbard dimer~\cite{10.1063/5.0291280}, the bulk silicon~\cite{2026arXiv260327329R} and the valence and core photoemission spectra of atoms and molecules~\cite{2026arXiv260720070P} where it was shown to successfully capture key features of both quasiparticles and satellites. However, several open questions remain such as the accuracy of the approximation in the large-system limit, its formal connection to existing approximation schemes, and crucially, its applicability to open-shell systems -- a class of problems of broad physical relevance.

Open-shell systems represent a particularly rich and challenging domain for the description of charged excitations. They arise in a wide range of physical contexts, e.g. reactive molecular environments, atomic nuclei, and magnetic or strongly correlated materials hosting localized spins and spin-polarized quasi-particles. Despite their ubiquity, a consistent and accurate theoretical treatment of charged excitations in open-shell systems remains an open challenge. 

Among approximation schemes available to tackle this challenge, the algebraic diagrammatic construction (ADC) scheme carries strong potential. The ADC~\cite{PhysRevA.26.2395,PhysRevA.43.4647,PhysRevA.53.2140,10.1063/1.1752875,PhysRevA.28.1237,https://doi.org/10.1002/wcms.1206} scheme delivers the so-called ADC(n) hierarchy of computationally efficient approximations of the one-body dynamical self energy. In particular, the non-Dyson ADC schemes~\cite{10.1063/1.477085,10.1063/1.2047550,10.1063/1.4931643,10.1063/1.5081674,10.1063/1.5131771}, such as ADC(2)-X, achieve an accuracy comparable to coupled-cluster theory with singles and doubles~\cite{doi:https://doi.org/10.1002/9780470125915.ch2,shavitt2009many} at a significantly reduced computational cost, when applied to charged excitations of molecules. The ADC has also been applied to closed-shell nuclei at the ADC(3) level~\cite{Raimondi:2017kzi,Raimondi:2017mey,Idini:2016oix} and to open-shell nuclei at the ADC(2) level~\cite{PhysRevC.105.044330,PhysRevC.84.064317,Soma:2013ona}. Such ADC-based GF calculations of open-shell molecular systems (atomic nuclei) rely on the use of a symmetry unrestricted Hartree-Fock(-Bogoliubov) reference state introducing spin (particle number) contamination~\cite{10.1063/5.0097333,doi:10.1021/acs.jctc.3c00251} (\cite{PhysRevC.105.044330,PhysRevC.84.064317,Soma:2013ona}).

Despite their independent development, the MCDE and ADC frameworks can both be recast as an eigenvalue problem involving an effective Hamiltonian. Such a structural similarity suggests the existence of a deeper formal connection between the two approximation schemes that needs to be uncovered. Thus, the present work establishes a formal equivalence between (3,1)-MCDE and an extension of Dyson ADC(2), analogous to the ADC(2)-X approximation~\cite{Trofimov1995,Barbieri:2016uib,10.1063/5.0097333,doi:10.1021/acs.jctc.3c00251}. This  rigorous connection between two independently developed approximation  schemes resolves  open questions regarding the accuracy of the MCDE in the large-system limit and its connection to pre-existing approximations. Furthermore, this equivalence immediately enables the extension of the (3,1)-MCDE to open-shell systems, by leveraging the existing ADC machinery developed for open-shell molecules~\cite{10.1063/5.0097333,doi:10.1021/acs.jctc.3c00251} and nuclei~\cite{PhysRevC.105.044330,PhysRevC.84.064317,Soma:2013ona} based on symmetry-unrestricted reference states.

The article is organized as follows. While Sec.~\ref{sec:scgf} introduces the general Green's function many-body framework, Secs.~\ref{sec:mcde} and~\ref{sec:adc} specify the (3,1)-MCDE and ADC(2/3) formalisms, respectively. In Sec.~\ref{sec:matching}, the formal equivalence between (3,1)-MCDE and an extension of ADC(2) is established. Finally, results of ADC(2) and (3,1)-MCDE truncation schemes for the  4-, 6-, and 8-site Hubbard rings are compared in Sec.~\ref{sec:Hubb}. Conclusions are provided in Sec.~\ref{conclusions} whereas a set of appendices deliver complementary details and perspectives.

\section{Many-body Problem}
\label{sec:ADCMCDE}

\subsection{Self-consistent Green's function theory}
\label{sec:scgf}

The interacting many-fermion system is governed by the Hamiltonian\footnote{Higher-rank interactions, such as mandatory three-nucleon forces in nuclear systems, are presently omitted for simplicity. Adding them makes the formalism and  associated numerical calculations more elaborate but does not pose any fundamental difficulty.}
\begin{comment}
\begin{align}
\label{eq:H}
H
&\equiv T + V \nonumber \\
&\equiv \sum_{\alpha\beta} t_{\alpha\beta}\, a^\dagger_\alpha a_\beta
+ \frac{1}{4}\sum_{\alpha\beta\gamma\delta} \bar{v}_{\alpha\beta\gamma\delta}\,
a^\dagger_\alpha a^\dagger_\beta a_\delta a_\gamma \, ,
\end{align}
\end{comment}

\begin{align}
\label{eq:H}
H
&\equiv T + V \nonumber \\
&\equiv \sum_{\alpha\beta} t_{\alpha\beta}\, a^\dagger_\alpha a_\beta
+ \frac{1}{4}\sum_{\alpha\beta\gamma\delta} \bar{v}_{\alpha\beta\gamma\delta}\,
a^\dagger_\alpha a^\dagger_\beta a_\gamma a_\delta\,,
\end{align}
where greek indices label a complete orthonormal one-body basis $\{| \alpha \rangle\}$ with associated creation (annihilation) operators $\{a^\dagger_\alpha\}$ ($\{a_\alpha\}$). The four-index tensor $\bar v_{\alpha\beta\gamma\delta}$ denotes antisymmetrized matrix elements of the two-body interaction

\begin{comment}
\begin{subequations}
\begin{align}    
    \bar v_{\alpha\beta\gamma\delta} &\equiv \langle\alpha\beta|v|\gamma\delta\rangle - \langle\alpha\beta|v|\delta\gamma\rangle\,,\\
    \qquad \langle\alpha\beta|v|\gamma\delta\rangle &\equiv \int dx_1 dx_2\,\varphi^*_\alpha(x_1)\varphi^*_\beta(x_2)\,v\,\varphi_\gamma(x_1)\varphi_\delta(x_2)\,,
\end{align}
\end{subequations}
\end{comment}
\begin{subequations}
\begin{align}
    \bar v_{\alpha\beta\gamma\delta} &\equiv v_{\alpha\beta\gamma\delta} - v_{\alpha\beta\delta\gamma}\,,\\
    v_{\alpha\beta\gamma\delta} &\equiv \int dx_1 dx_2\,
    \varphi^*_\alpha(x_1)\varphi^*_\beta(x_2)\,v(x_1,x_2)\,
    \varphi_\gamma(x_2)\varphi_\delta(x_1)\,,
\end{align}
\end{subequations}
whereas $t_{\alpha\beta}$ represents matrix elements of the one-body kinetic energy term. 

Denoting by $\ket{\Psi_0^N}$ the $N$-body ground state of $H$ with energy $E_0^N$, the time-ordered, zero temperature and equilibrium one-body Green's function is defined as
\begin{equation}
    ig_{\alpha\beta}(t_1,t_2) \equiv \bra{\Psi_0^N}\mathcal{T}\{a_\alpha(t_1)a_\beta^\dagger(t_2)\}\ket{\Psi_0^N}\,,
\end{equation}
with $\mathcal{T}$ the time-ordering operator and $\{a^\dagger_\alpha(t_1)\}$ ($\{a_\alpha(t_1)\}$) the creation (annihilation) operators in the Heisenberg representation. Going to the energy domain, the one-body Green's function satisfies the Lehmann representation
\begin{equation}
\label{eq:1BGF}
g_{\alpha\beta}(\omega)
= \sum_n \frac{\mathcal{X}^{n*}_\alpha \mathcal{X}^n_\beta}{\omega-\varepsilon^+_n+i\eta}
+ \sum_m \frac{\mathcal{Y}^m_\alpha \mathcal{Y}^{m*}_\beta}{\omega-\varepsilon^-_m-i\eta}\, ,
\end{equation}
where $n$ ($m$) runs over exact eigenstates of the $(N\!+\!1)$-body ($(N\!-\!1)$-body) systems. Here, $\omega$ is the energy variable and $\eta\to0^+$ is a positive infinitesimal that sets the time-ordering of the 1-GF, placing the addition (removal) poles infinitesimally below (above) the real axis. The spectroscopic amplitudes and separation energies associated with one-fermion addition and removal processes are defined as
\begin{subequations}
\begin{align}
\mathcal{X}^n_\beta &\equiv \bra{\Psi_n^{N+1}} a^\dagger_\beta \ket{\Psi_0^N}\, , \\
\mathcal{Y}^m_\alpha &\equiv \bra{\Psi_m^{N-1}} a_\alpha \ket{\Psi_0^N}\, , \\
\varepsilon^+_n &\equiv E_n^{N+1}-E_0^N\, , \\
\varepsilon^-_m &\equiv E_0^N-E_m^{N-1}\, .
\end{align}
\end{subequations}
Thus, $\varepsilon^+_n$ ($\varepsilon^-_m$) corresponds to the energy required to add (remove) one particle to (from) the $N$-body ground state, leaving the system in the final state
$\ket{\Psi_n^{N+1}}$ ($\ket{\Psi_m^{N-1}}$).

The interacting 1-GF of Eq.~\eqref{eq:1BGF} satisfies Dyson's equation~\cite{PhysRev.75.1736}
\begin{equation}
\label{eq:1BDE}
g_{\tau\iota}(\omega)
= g^{(0)}_{\tau\iota}(\omega)
+ \sum_{\alpha\beta} g^{(0)}_{\tau\alpha}(\omega)\,\Sigma^*_{\alpha\beta}(\omega)\,g_{\beta\iota}(\omega)\, ,
\end{equation}
where $g^{(0)}(\omega)$ is the non-interacting one-body Green's function and $\Sigma^*(\omega)$ the one-particle irreducible self-energy. The latter can be partitioned according to~\cite{Barbieri:2016uib}
\begin{equation}
\label{eq:self-energy}
\Sigma_{\alpha\beta}^*(\omega) \equiv \Sigma_{\alpha\beta}^{(\infty)}+\tilde{\Sigma}_{\alpha\beta}(\omega)\, ,
\end{equation}
where $\Sigma^{(\infty)}$ constitutes the $\omega$-independent static contribution whereas $\tilde{\Sigma}(\omega)$ accounts for dynamical correlations. 

In Eq.~\eqref{eq:1BDE}, the interacting 1-GF $g(\omega)$ is expanded with respect to a non-interacting 1-GF  $g^{(0)}(\omega)$. The expansion can however be re-written in terms of a reference Green's function associated with a Slater determinant that already contains information about the interactions  between the particles in a mean-field approximation. In most applications, one typically starts from $g^{(\mathrm{HF})}(\omega)$ associated with a Hartree-Fock (HF) reference Slater determinant. Incorporating a certain degree of self-consistency in the Green's function calculation, one can rather employ the reference propagator $g^{(\infty)}$ satisfying
\begin{equation}
\label{eq:ginf_def}
g^{(\infty)}_{\tau\iota}(\omega)
=
g^{0}_{\tau\iota}(\omega)
+\sum_{\alpha\beta}g^{0}_{\tau\alpha}(\omega)\,\Sigma^{(\infty)}_{\alpha\beta}\,g^{(\infty)}_{\beta\iota}(\omega)\, ,
\end{equation}
such that Dyson's equation is rewritten as
\begin{equation}
\label{eq:1BDEinf}
g_{\tau\iota}(\omega)
=
g^{(\infty)}_{\tau\iota}(\omega)
+\sum_{\alpha\beta}g^{(\infty)}_{\tau\alpha}(\omega)\,\tilde{\Sigma}_{\alpha\beta}(\omega)\,g_{\beta\iota}(\omega)\, .
\end{equation}

Either way, the reference propagator relates to a Slater determinant built out of the eigenstates  $\{| \alpha \rangle\}$ of a one-body Hamiltonian of choice, e.g. the Hartree-Fock one-body Hamiltonian $h^{(\mathrm{HF})}\equiv T + U^{\mathrm{HF}}$ or the so-called Baranger one-body Hamiltonian~\cite{Baranger:1970abt} $h^{(\infty)}\equiv T + \Sigma^{(\infty)}$. The static one-body potential $\Sigma^{(\infty)}$ is nothing but a {\it correlated} HF potential in which the two-body interaction is convoluted with the fully-correlated one-body density matrix rather than with the Slater determinant one-body density matrix~\cite{Baranger:1970abt,Duguet:2011sq}. Denoting generically by $\{\varepsilon_\alpha\}$ the eigenvalues of $h^{(\infty/\mathrm{HF})}$, the reference propagator takes the generic form
\begin{equation}
\label{eq:HF1BGF}
\begin{aligned}
g_{\alpha\beta}^{(\infty/\mathrm{HF})}(\omega)
=& \delta_{\alpha\beta}\left[
\frac{1-f_\alpha}{\omega-\varepsilon_\alpha+i\eta}
+\frac{f_\alpha}{\omega-\varepsilon_\alpha-i\eta}
\right]\\
=&\frac{\delta_{\alpha\beta}}{\omega-\varepsilon_\alpha+i\eta\,\mathrm{sign}(\mu-\varepsilon_\alpha)}\, ,
\end{aligned}
\end{equation}
where $f_\alpha\in\{0,1\}$ denotes the occupation of the single-particle state $| \alpha \rangle$ in the reference Slater determinant.

While the chosen reference state does not impact the results in a {\it self-consistent} setting of Green's function theory where the self-energy is itself expanded in terms of the fully dressed propagator, it does have an impact when the self energy is rather expanded in terms of that reference state propagator as is presently considered~\cite{10.1063/5.0291280}.

Within the Green's-function formalism briefly introduced above, the working approximation resides in the definition of $\tilde{\Sigma}(\omega)$. The MCDE and ADC expansion schemes are introduced in the following two sections at the (3,1)-MCDE and ADC(2/3) level, respectively.

\subsection{Multichannel Dyson Equation}
\label{sec:mcde}

Based on a HF reference state, the (3,1)-MCDE expansion scheme~\cite{PhysRevLett.131.216401,PhysRevB.110.115140} couples the one-body propagator to the $2e1h$ and $1e2h$ channels of the three-body Green's function. The dynamical equation to solve can be written compactly as a block Dyson equation of the form
% --- à placer dans le préambule ---
\begin{subequations}
\label{eq:MCDE}
\setlength{\arraycolsep}{3pt}
\renewcommand{\arraystretch}{1.2}
\begin{align}
  \mathbf{G} &= \mathbf{G}^{0} + \mathbf{G}^{0}\, \boldsymbol{\Sigma}\, \mathbf{G} \, ,
  \label{eq:basicMCDEequation} \\
  \intertext{with the block matrices}
  \mathbf{G} &=
  \begin{pmatrix}
    g_{\alpha\beta} & G^{1p/2e1h}_{\alpha r'} & G^{1p/1e2h}_{\alpha s'} \\
    {G}^{2e1h/1p}_{r\beta} & \mathcal{G}^{2e1h}_{rr'} & 0 \\
    {G}^{1e2h/1p}_{s\beta} & 0 & \mathcal{G}^{1e2h}_{ss'}
  \end{pmatrix} ,
  \label{eq:MCDEblockG} \\[2ex]
  \mathbf{G}^{0} &=
  \begin{pmatrix}
    g^{\mathrm{HF}}_{\alpha\beta} & 0 & 0 \\
    0 & G^{0,2e1h}_{rr'} & 0 \\
    0 & 0 & G^{0,1e2h}_{ss'}
  \end{pmatrix} ,
  \label{eq:MCDEblockG0} \\[2ex]
   \boldsymbol{\Sigma} &=
  \begin{pmatrix}
    0 & \Sigma^{\mathrm{1p}/2e1h}_{\alpha r'} & \Sigma^{\mathrm{1p}/1e2h}_{\alpha s'} \\
    \Sigma^{2e1h/\mathrm{1p}}_{r\beta} & \Sigma^{2e1h}_{rr'} & 0 \\
    \Sigma^{1e2h/\mathrm{1p}}_{s\beta} & 0 & \Sigma^{1e2h}_{ss'}
  \end{pmatrix} ,
  \label{eq:MCDEblockSig}
\end{align}
\end{subequations}
and where $r$ and $s$ denote {\it composite} indices labeling intermediate-state {$2e1h$} and {$1e2h$} configurations (ISCs), respectively,
\begin{equation}
r \equiv (n_1 n_2 k_3)\, , \qquad s \equiv (k_1 k_2 n_3)\, ,
\end{equation}
where unoccupied ({electron}) and occupied (hole) single-particle states in the reference Slater determinant are denoted by $n$ and $k$, respectively, and satisfy $f_n=0$  and $f_k=1$. Unperturbed ISC energies are collected in diagonal matrices $E^{>}$ and
$E^{<}$, i.e.\ $E^{>}_{rr'}=E^{>}_{r}\delta_{rr'}$ and $E^{<}_{ss'}=E^{<}_{s}\delta_{ss'}$, with
\begin{subequations}
\label{eq:energy_2e1h_2h1p}
\begin{align}
E^{>}_{r} \equiv& \varepsilon_{n_1}+\varepsilon_{n_2}-\varepsilon_{k_3}\, ,\\
E^{<}_{s} \equiv& \varepsilon_{k_1}+\varepsilon_{k_2}-\varepsilon_{n_3}\, .
\end{align}
\end{subequations}
In Eq.~\eqref{eq:basicMCDEequation}, $\mathcal{G}^{2e1h}_{rr'}$ and $\mathcal{G}^{1e2h}_{ss'}$ denote the dressed {$2e1h$} and {$1e2h$}
propagators, respectively, whereas their reference counterparts read as
\begin{subequations}
\begin{align}
    G^{0,2e1h}_{rr'}\equiv & \frac{\delta_{rr'}}{\omega-E_r^>+i\eta}\,,\\
    G^{0,1e2h}_{ss'}\equiv & \frac{\delta_{ss'}}{\omega-E_s^<-i\eta}\,.
\end{align}
\end{subequations}
The off-diagonal blocks $G_{\alpha r'}^{1p/2e1h}$ ($G_{r\beta}^{2e1h/1p}$) and ${G}^{1p/1e2h}_{\alpha s'}$ (${G}^{1e2h/1p}_{s\beta}$) are the propagators coupling the one-body and three-body sectors.

Within the (3,1)-MCDE approximation, the multichannel self-energy entering the block self-energy reads as
\begin{comment}
\begin{subequations}
\label{eq:selfen3}
\begin{align}
    \Sigma^c_{\alpha r} \equiv& \bar v_{\alpha k_3 n_1 n_2}\,,\\
    \Sigma^c_{\alpha s} \equiv& \bar v_{\alpha n_3 k_1 k_2}\,,\\
    \tilde\Sigma^c_{r\beta} \equiv& \bar v_{n_1 n_2 \beta k_3}\,,\\
    \tilde\Sigma^c_{s\beta} \equiv& \bar v_{k_1 k_2 \beta n_3}\,,\\
    \Sigma^{3p}_{rr'} \equiv & \bar{v}_{n_1n_2n_1'n_2'}\delta_{k_3k_3'}
    -\bar{v}_{n_1k_3'k_3n_2'}\delta_{n_2n_1'}
    -\bar{v}_{n_2k_3'k_3n_1'}\delta_{n_1n_2'}\notag \\
    &+\bar{v}_{n_1k_3'k_3n_1'}\delta_{n_2n_2'}
    +\bar{v}_{n_2k_3'k_3n_2'}\delta_{n_1n_1'}\,,\\
    \Sigma^{3p}_{ss'} \equiv & -\bar{v}_{k_1k_2k_1'k_2'}\delta_{n_3n_3'}
    +\bar{v}_{k_1n_3'n_3k_2'}\delta_{k_2k_1'}
    +\bar{v}_{k_2n_3'n_3k_1'}\delta_{k_1k_2'}\notag \\
    &-\bar{v}_{k_1n_3'n_3k_1'}\delta_{k_2k_2'}
    -\bar{v}_{k_2n_3'n_3k_2'}\delta_{k_1k_1'}\,.
\end{align}
\end{subequations}
\end{comment}
\begin{subequations}
\label{eq:selfen3}
\begin{align}
    \Sigma^{\text{1p}/{2e1h}}_{\alpha r} \equiv& \bar v_{\alpha k_3 n_2 n_1}\,,\\
    \Sigma^{\text{1p}/1e2h}_{\alpha s} \equiv& \bar v_{\alpha n_3 k_2 k_1}\,,\\
    \Sigma^{2e1h/\text{1p}}_{r\beta} \equiv& \bar v_{n_1 n_2 k_3 \beta}\,,\\
    \Sigma^{1e2h/\text{1p}}_{s\beta} \equiv& \bar v_{k_1 k_2 n_3 \beta}\,,\\
    \Sigma^{2e1h}_{rr'} \equiv & \bar{v}_{n_1 n_2 n_2' n_1'}\,\delta_{k_3 k_3'}
    +\bar{v}_{n_1 k_3' k_3 n_2'}\,\delta_{n_2 n_1'}
    +\bar{v}_{n_2 k_3' k_3 n_1'}\,\delta_{n_1 n_2'}\notag \\
    &-\bar{v}_{n_1 k_3' k_3 n_1'}\,\delta_{n_2 n_2'}
    -\bar{v}_{n_2 k_3' k_3 n_2'}\,\delta_{n_1 n_1'}\,,\\
    \Sigma^{1e2h}_{ss'} \equiv & -\bar{v}_{k_1 k_2 k_2' k_1'}\,\delta_{n_3 n_3'}
    -\bar{v}_{k_1 n_3' n_3 k_2'}\,\delta_{k_2 k_1'}
    -\bar{v}_{k_2 n_3' n_3 k_1'}\,\delta_{k_1 k_2'}\notag \\
    &+\bar{v}_{k_1 n_3' n_3 k_1'}\,\delta_{k_2 k_2'}
    +\bar{v}_{k_2 n_3' n_3 k_2'}\,\delta_{k_1 k_1'}\,.
\end{align}
\end{subequations}

% \begin{subequations}
% \label{eq:MCDEcoup}
% \begin{align}
% \Sigma^c_{\alpha(\gamma\delta\sigma)} \equiv& \bar{v}_{\alpha\sigma\gamma\delta}\, ,\\
% \tilde{\Sigma}^c_{(\theta\zeta\rho)\beta} \equiv& \bar{v}_{\theta\zeta\beta\rho}\, ,\\
% \Sigma^{3p}_{(\phi\rho\psi)(\theta\zeta\chi)} \equiv&
% \Big[(1-n_\phi)(1-n_\rho)n_\psi-n_\phi n_\rho(1-n_\psi)\Big]\notag\\
% &\times\Big[\bar{v}_{\phi\rho\theta\zeta}\delta_{\psi\chi}
% +\bar{v}_{\phi\chi\psi\zeta}\delta_{\rho\theta}
% +\bar{v}_{\rho\chi\psi\theta}\delta_{\phi\zeta}
% \\
% &-\bar{v}_{\phi\chi\psi\theta}\delta_{\rho\zeta}
% -\bar{v}_{\rho\chi\psi\zeta}\delta_{\phi\theta}\Big]\notag\ .
% \end{align}
% \end{subequations}

Solving Eq.~\eqref{eq:basicMCDEequation} yields $g(\omega)$ as an approximate solution of Dyson's equation corresponding to an implicit approximation to the dynamical self-energy $\tilde{\Sigma}(\omega)$ induced by the explicit coupling to the three-body sector. It is an objective of the present work to {further} clarify the content of such an approximation to $\tilde{\Sigma}(\omega)$. 

Introducing the  energy-independent effective-Hamiltonian
\begin{equation}
\label{eq:Heff_MCDE}
\mathcal{H}^{\rm (3,1)\text{-}MCDE}
=
\begin{pmatrix}
H^{\rm 1p}_{\alpha\beta} & \Sigma^{\text{1p}/2e1h}_{\alpha r} & \Sigma^{\text{1p}/1e2h}_{\alpha s}  \\
\Sigma^{2e1h/\text{1p}}_{r\beta} & H^{2e1h}_{rr'} & 0\\
\Sigma^{1e2h/\text{1p}}_{s\beta} & 0 & H^{1e2h}_{ss'}\\
\end{pmatrix}\, ,
\end{equation}
with
\begin{subequations}
    \begin{equation}
        H^{\rm 1p}_{\alpha\beta} \equiv \varepsilon_\alpha\,\delta_{\alpha\beta}\,,
    \end{equation}
    \begin{equation}
        H^{ 2e1h}_{rr'} \equiv E^>_r\delta_{rr'}+\Sigma_{rr'}^{2e1h}\,,
    \end{equation}
    \begin{equation}
        H^{ 1e2h}_{ss'} \equiv E^<_s\delta_{ss'}+\Sigma_{ss'}^{1e2h}\, ,
    \end{equation}
\end{subequations}
% \begin{subequations}
% \begin{align}
% H^{1p}_{\alpha\beta} \equiv& \varepsilon_\alpha\,\delta_{\alpha\beta}\, ,\\
% H^{3p}_{(\alpha\beta\gamma)(\mu\nu\lambda)} \equiv&
% (\varepsilon_\alpha+\varepsilon_\beta-\varepsilon_\gamma)\,
% \delta_{\alpha\mu}\delta_{\beta\nu}\delta_{\gamma\lambda}
% \\
% &+\Sigma^{3p}_{(\alpha\beta\gamma)(\mu\nu\lambda)}\notag\, .
% \end{align}
% \end{subequations}
Eq.~\eqref{eq:basicMCDEequation} can be recast as the diagonalization of $\mathcal{H}^{\rm (3,1)\text{-}MCDE}$. Denoting as $E_\lambda$ ($A_\lambda$) the corresponding eigenvalues (eigenvectors), the one-body propagator can be finally retrieved as
\begin{equation}
\label{eq:MCDEGREEN}
g_{\alpha\beta}(\omega)
=
\sum_\lambda
\frac{A^\alpha_\lambda\,(A^\beta_\lambda)^{*}}{\omega-E_\lambda}\, 
\end{equation}

This effective Hamiltonian representation will be used to connect the (3,1)-MCDE truncation to  a specific approximation obtained within the ADC expansion scheme that is now briefly introduced.

\subsection{Algebraic Diagrammatic Construction}
\label{sec:adc}

ADC provides a systematic hierarchy of approximations to the dynamical self-energy $\tilde{\Sigma}(\omega)$ entering Eq.~\eqref{eq:self-energy}, which can be expressed in the Lehmann representation of the form~\cite{PhysRevA.28.1237}
\begin{equation}
\label{eq:lehmann}
\begin{aligned}
\tilde{\Sigma}_{\alpha\beta}(\omega)
\equiv&
\sum_{rr'} \mathcal{M}^\dagger_{\alpha r}
\left[\frac{1}{\omega-(E^{>}+\mathcal{C})+i\eta}\right]_{rr'}
\mathcal{M}_{r'\beta}
\\
&+
\sum_{ss'} \mathcal{N}_{\alpha s}
\left[\frac{1}{\omega-(E^{<}+\mathcal{D})-i\eta}\right]_{ss'}
\mathcal{N}^\dagger_{s'\beta}\, .
\end{aligned}
\end{equation}
By construction this form guarantees a positive-definite spectral function~\cite{Stefanucci_2014,Barbieri:2016uib}. While matrices $\mathcal{M}$ and $\mathcal{N}$ couple one-body states to {$2e1h$} and {$1e2h$} ISCs, respectively, $\mathcal{C}$ and $\mathcal{D}$ describe interactions within each of these two ISC spaces. 

The ADC procedure constructs the coupling and interaction matrices by matching an order-by-order perturbative expansion of the self-energy. Starting from a formal perturbative expansion
\begin{equation}
\label{eq:adc_series_example}
\mathcal{M}_{\alpha r}
=\mathcal{M}^{(1)}_{\alpha r}+\mathcal{M}^{(2)}_{\alpha r}+\cdots\, ,
\end{equation}
and analogously for $\mathcal{N}$, $\mathcal{C}$, and $\mathcal{D}$, the resolvent expansion
\begin{equation}
\label{eq:resolvent_series}
(P-\mathcal{Q})^{-1}
= P^{-1}+P^{-1}\mathcal{Q}P^{-1}+P^{-1}\mathcal{Q}P^{-1}\mathcal{Q}P^{-1}+\cdots\, ,
\end{equation}
with $\mathcal{Q}\in\{\mathcal{C},\mathcal{D}\}$ and $P\equiv \omega-{E^{\substack{<\\>}}}\pm i\eta$, is used to expand Eq.~\eqref{eq:lehmann}. Then, matrices $\mathcal{M}^{(k)}$, $\mathcal{N}^{(k)}$,
$\mathcal{C}^{(k)}$, and $\mathcal{D}^{(k)}$ are determined by matching this series to the perturbative expansion of the self-energy computed at a chosen order $n$. Inserting the resulting matrices
back into Eq.~\eqref{eq:lehmann} defines the ADC($n$) approximation that reproduces by construction all perturbative contributions to $\tilde\Sigma_{\alpha\beta}(\omega)$ through order $n$ while retaining the spectral (Lehmann) form that effectively corresponds to resumming selected higher-order terms.

For $n=2$, the ADC(2) approximation reads as
\begin{equation}
\label{eq:order2}
\begin{aligned}
\tilde\Sigma^{\mathrm{ADC(2)}}_{\alpha\beta}(\omega)
=&
\sum_{r}\mathcal{M}^{(1)\dagger}_{\alpha r}
\left[\frac{1}{\omega-E_r^>+i\eta}\right]\mathcal{M}^{(1)}_{r \beta}
\\
&+\sum_{s}\mathcal{N}^{(1)}_{\alpha s}
\left[\frac{1}{\omega-E_s^<-i\eta}\right]\mathcal{N}^{(1)\dagger}_{s \beta}\, .
\end{aligned}
\end{equation}

For $n=3$, the ADC(3) approximation involves (i) first-order interaction matrices $\mathcal{C}^{(1)}$ and $\mathcal{D}^{(1)}$ and (ii) second-order corrections to the coupling matrices, i.e. $\mathcal{M}^{(1,2)}=\mathcal{M}^{(1)}+\mathcal{M}^{(2)}$ and $\mathcal{N}^{(1,2)}=\mathcal{N}^{(1)}+\mathcal{N}^{(2)}$, such that
\begin{equation}
\label{eq:ADC3_Lehmann}
\begin{aligned}
\tilde\Sigma^{\mathrm{ADC(3)}}_{\alpha\beta}(\omega)
=&\sum_{rr'}
\mathcal{M}^{(1,2)^\dagger}_{\alpha r}
\left[\frac{1}{\omega-(E^>+\mathcal{C}^{(1)})+i\eta}\right]_{rr'}
\mathcal{M}^{(1,2)}_{r'\beta}\\
&+\sum_{ss'}
\mathcal{N}^{(1,2)}_{\alpha s}
\left[\frac{1}{\omega-(E^<+\mathcal{D}^{(1)})-i\eta}\right]_{ss'}
\mathcal{N}^{(1,2)\dagger}_{s'\beta}\, .
\end{aligned}
\end{equation}

Eventually, solving Dyson's equation based on the ADC(n) approximation can be recast as the
diagonalization of an energy-independent effective Hamiltonian, e.g.
%\begin{widetext}
\begin{equation}
\label{eq:Heff_ADC2}
\mathcal{H}^{\mathrm{ADC(2)}}
=
\begin{pmatrix}
T+\Sigma^{(\infty)} &
\mathcal{M}^{(1)\dagger} &
\mathcal{N}^{(1)}\\
\mathcal{M}^{(1)} &
E^>  & 0\\
\mathcal{N}^{(1)\dagger}&
0 & E^<
\end{pmatrix}\, ,
\end{equation}
%\end{widetext}
and
\begin{equation}
\label{eq:Heff_ADC3}
\mathcal{H}^{\mathrm{ADC(3)}}
=
\begin{pmatrix}
T+\Sigma^{(\infty)} &
\mathcal{M}^{(1,2)\dagger} &
\mathcal{N}^{(1,2)}\\
\mathcal{M}^{(1,2)} &
E^> + \mathcal{C}^{(1)}  & 0\\
\mathcal{N}^{(1,2)\dagger}&
0 & E^< +\mathcal{D}^{(1)}
\end{pmatrix}\, ,
\end{equation}

\subsection{Connection between the two schemes}
\label{sec:matching}

Both ADC and MCDE expansions admit an effective-Hamiltonian formulation. Employing the HF reference state in the ADC approach as is customary in MCDE, the two schemes can now be connected by comparing the building blocks of $\mathcal{H}^{\mathrm{ADC(2)}}$ and $\mathcal{H}^{\mathrm{ADC(3)}}$ to those of {$\mathcal{H}^{(3,1)-\mathrm{MCDE}}$} that have been introduced explicitly in Sec.~\ref{sec:mcde}. Note that an alternative route to establish this connection and derive the (3,1)-MCDE is followed in App.~\ref{sec:adc-mcde}.

Following the ADC(n) {rationale} for $n=2$ and $n=3$~\cite{PhysRevA.28.1237}, the generic explicit form of $\mathcal{M}^{(1,2)}$ and $\mathcal{N}^{(1,2)}$, as well as of $\mathcal{C}^{(1)}$ and $\mathcal{D}^{(1)}$, in terms of two-body interaction matrix elements and spectroscopic amplitudes is given in App.~\ref{app:adc_matchings}. Specifying them to the HF reference state, one obtains the strict correspondences
\begin{subequations}
\label{connectionmat}
    \begin{equation}
        h^{(\mathrm{HF})}_{\alpha\beta} = H^{\rm 1p}_{\alpha\beta} \, ,
    \end{equation}
    \begin{equation}
        \mathcal{M}_{\alpha r}^{(1)\dagger} = \Sigma^{\text{1p}/2e1h}_{\alpha r} \, ,
    \end{equation}
    \begin{equation}
        \mathcal{N}^{(1)}_{\alpha s} = \Sigma^{\text{1p}/1e2h}_{\alpha s} \, ,
    \end{equation}
    \begin{equation}
        \mathcal{C}^{(1)}_{rr'} = \Sigma^{2e1h}_{rr'} \, ,
    \end{equation}
    \begin{equation}
        \mathcal{D}^{(1)}_{ss'} = \Sigma^{1e2h}_{ss'} \, ,
    \end{equation}
\end{subequations}
leading to

\begin{widetext}
\begin{equation}
\label{eq:Heff_ADC_HF}
\setlength{\arraycolsep}{2pt}
\renewcommand{\arraystretch}{1.3}
\mathcal{H}^{\mathrm{ADC(2)}}
=
\begin{pmatrix}
  H^{\mathrm{1p}}_{\alpha\beta} &
  \Sigma^{\mathrm{1p}/2e1h}_{\alpha r'} &
  \Sigma^{\mathrm{1p}/1e2h}_{\alpha s'} \\
  \Sigma^{2e1h/\mathrm{1p}}_{r\beta} &
  E^{>}_{r}\delta_{rr'} & 0 \\
  \Sigma^{1e2h/\mathrm{1p}}_{s\beta} &
  0 & E^{<}_{s}\delta_{ss'}
\end{pmatrix} ,
\qquad
\mathcal{H}^{\mathrm{ADC(3)}}
=
\begin{pmatrix}
  H^{\mathrm{1p}}_{\alpha\beta} &
  \Sigma^{\mathrm{1p}/2e1h}_{\alpha r'} + \mathcal{M}^{(2)\dagger}_{\alpha r'} &
  \Sigma^{\mathrm{1p}/1e2h}_{\alpha s'} + \mathcal{N}^{(2)}_{\alpha s'} \\
  \Sigma^{2e1h/\mathrm{1p}}_{r\beta} + \mathcal{M}^{(2)}_{r\beta} &
  H^{2e1h}_{rr'} & 0 \\
  \Sigma^{1e2h/\mathrm{1p}}_{s\beta} + \mathcal{N}^{(2)\dagger}_{s\beta} &
  0 & H^{1e2h}_{ss'}
\end{pmatrix} .
\end{equation}
\end{widetext}

The (3,1)-MCDE effective Hamiltonian defined in Eq.~\eqref{eq:Heff_MCDE} can be obtained either by adding the interaction matrices $\Sigma^{2e1h}$ and $\Sigma^{1e2h}$ to the diagonal blocks of $\mathcal{H}^{\mathrm{ADC(2)}}$ or by removing the second-order coupling matrices from the off-diagonal blocks of $\mathcal{H}^{\mathrm{ADC(3)}}$.
It follows that the self-energy corresponding to the (3,1)-MCDE approximation lies in between ADC(2) and ADC(3) truncations and can be written as %~\cite{2026arXiv260327329R} \PR{I put the reference to Arjan's paper here}
\begin{widetext}
\begin{equation}
\label{eq:MCDESE_11}
\begin{aligned}
\tilde\Sigma^{\rm (3,1)\text{-}MCDE}_{\alpha\beta}(\omega)
=& \sum_{rr'}\Sigma^{\text{1p}/2e1h}_{\alpha r}
\left[\frac{1}{\omega-(E^>+\Sigma^{2e1h})+i\eta}\right]_{rr'}
\Sigma^{2e1h/\rm 1p}_{r' \beta}+\sum_{ss'}\Sigma^{\text{1p}/1e2h}_{\alpha s}
\left[\frac{1}{\omega-(E^<+\Sigma^{1e2h})-i\eta}\right]_{ss'}
\Sigma^{1e2h/\rm 1p}_{s'\beta}\, ,
\end{aligned}
\end{equation}
\end{widetext}
which corresponds to the ADC(2)-X extension of ADC(2)~\cite{Trofimov1995,Barbieri:2016uib,10.1063/5.0097333,doi:10.1021/acs.jctc.3c00251}. {We note that this expression for $\tilde\Sigma^{\rm (3,1)\text{-}MCDE}$ was also reported in Ref.~\onlinecite{2026arXiv260327329R}.} 

\section{Application to the Hubbard model}
\label{sec:Hubb}

In light of the connection established between the (3,1)-MCDE approximation and the ADC(n) expansion scheme, (3,1)-MCDE and ADC(2) results are compared using the one-dimensional Hubbard model. To date, the application of the (3,1)-MCDE to the Hubbard model has so far been limited to the dimer case~\cite{PhysRevLett.131.216401,10.1063/5.0291280}. Here we extend its application to a larger number of sites.

The Hubbard Hamiltonian is a model Hamiltonian describing interacting fermions on a lattice. It consists of a kinetic hopping competing with a purely on-site two-body interaction. In this work we consider the model with only one orbital per site. The system presents a self-energy which displays non-trivial patterns with a strong frequency dependence producing a significant spectral fragmentation. 

The Hamiltonian reads as
\begin{equation}
\label{eq:Hubbard_H}
\begin{aligned}
\mathcal{H} 
=& -t\sum_{i=1}^{L}\sum_{\sigma=\uparrow,\downarrow} 
\Big(c_{i\sigma}^\dagger c_{(i+1)\sigma} + c_{(i+1)\sigma}^\dagger c_{i\sigma}\Big) \\
& + U\sum_{i=1}^{L} n_{i\uparrow} n_{i\downarrow} 
+ \varepsilon_0 \sum_{i=1}^{L}\sum_{\sigma=\uparrow,\downarrow} n_{i\sigma},
\end{aligned}
\end{equation}
where $c^\dagger_{i\sigma}$ ($c_{i\sigma}$) is the creation (annihilation) operator for orbital $i = 1,2,\dots,L$ with spin $\sigma = \uparrow,\downarrow$, $n_{i\sigma} = c^\dagger_{i\sigma} c_{i\sigma}$, $\varepsilon_0$ is the on-site energy, $t$ is the hopping parameter, and $U$ is the on-site Coulomb interaction.  

Calculations are presently performed {using periodic boundary conditions} at half-filling for system sizes $L \in \{4,6,8\}$, for which exact solutions can be computed numerically. 

\subsection{Restricted and Unrestricted HF reference state}
\label{sec:unres}

The (3,1)-MCDE and ADC(2) calculations are presently based on a nondegenerate HF reference state. The Hubbard ring model, at the HF level, exhibits a spontaneous breaking of spin symmetry within the mean-field approximation, resulting in a nonmagnetic-to-antiferromagnetic (NM-to-AFM) transition that does not occur in the exact ground state. This transition corresponds to the (spin-symmetry-broken) unrestricted HF (U-HF) solution becoming energetically favorable over the (spin-symmetry-preserving) restricted HF (R-HF) solution. While this spontaneous symmetry breaking occurs beyond a critical interaction strength in closed-shell systems, it occurs in open-shell systems as soon as the interaction strength is different from zero.

An efficient way to account for this feature in closed-shell systems is to employ a U-HF reference state in the strong-coupling regime. In open-shell systems, the use of a U-HF reference state is a convenient choice as soon as the interaction is nonzero. Indeed, a R-HF mean-field approximation, preserving spin symmetry, provides a rather poor description of the ground-state energy above the NM-to-AFM transition \cite{PhysRevB.111.125148,2022CoPP...62E0220J}, whereas U-HF captures the qualitative trend. In addition, for open-shell systems the R-HF reference state is degenerate at half-filling, unlike its U-HF counterpart.

\begin{figure}[t]
    \centering
    \includegraphics[width=0.9\linewidth]{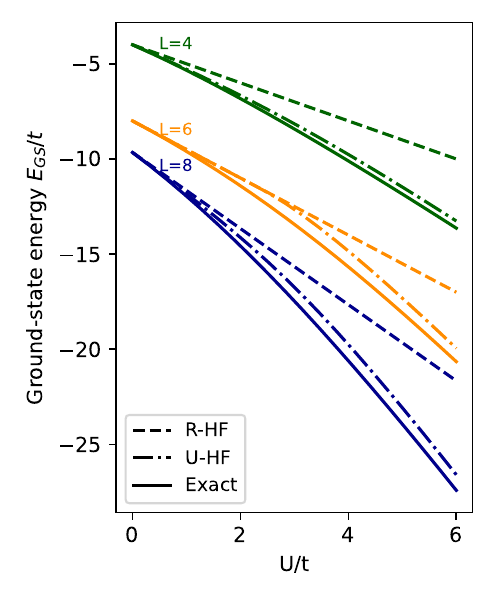}
    \caption{Exact versus R-HF and U-HF ground-state energies  as a function of $U/t$ for  $L=4,6,8$. A sharp phase transition {(i.e., the U-HF total energy (spin-broken solution) becomes lower than the R-HF one (spin-symmetric solution))} occurs at $U/t\approx 2.4$ for $L=6$ whereas it occurs as soon as U/t is different from zero for $L=4,8$. }
    \label{fig:phasetrans}
\end{figure}

These considerations are illustrated in Fig.~\ref{fig:phasetrans} for $L \in \{4,6,8\}$ at half-filling. The transition from NM to AFM occurs at $U/t \approx 2.4$ for the closed-shell case ($L=6$), whereas symmetry breaking sets in from $U=0$ (no finite critical coupling) for the open-shell systems ($L=4,8$)\footnote{As discussed in App.~\ref{sec:open}, the ground state of the one-dimensional Hubbard model with periodic boundary conditions is of open-shell character for $L=4M$ with $M\geq1$.}. In all cases, only the U-HF is able to describe qualitatively well the ground-state energy evolution over the full range of coupling strength $U/t$. 

The above observation, together with the need to employ a well-defined, i.e. nondegenerate, reference state with respect to particle-hole excitations to perform controlled GF calculations, motivates the use of the U-HF reference state, at the cost of breaking spin symmetry. In the following, calculations are performed for $U/t = 4$, which is a coupling strength beyond the mean-field NM–AFM transition even for $L=6$. {Solutions based on the U-HF (R-HF) reference state are referred to as U-ADC(2) (R-ADC(2)) and U-(3,1)-MCDE (R-(3,1)-MCDE)}.

\subsection{Spectral functions}

\begin{figure*}
    \centering
    \includegraphics[width=0.47\linewidth]{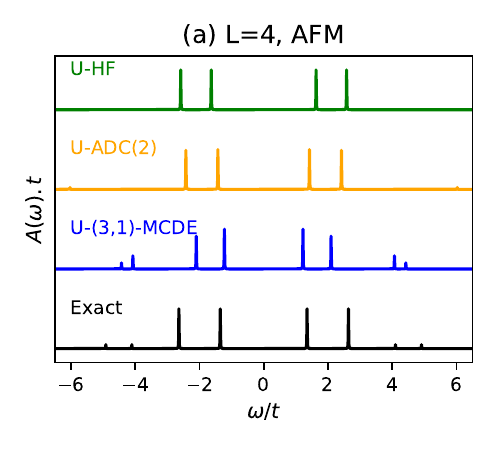}
    \vspace{0.02\linewidth}
    \includegraphics[width=0.47\linewidth]{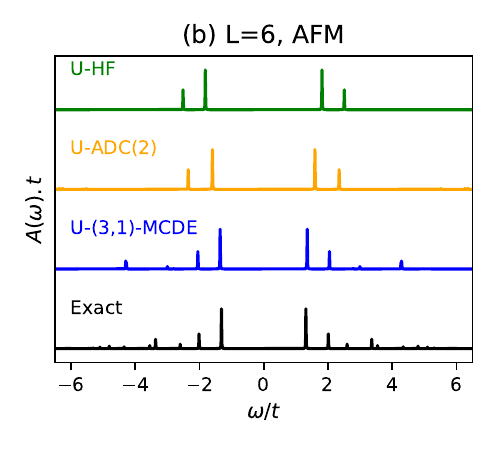}
    \includegraphics[width=0.47\linewidth]{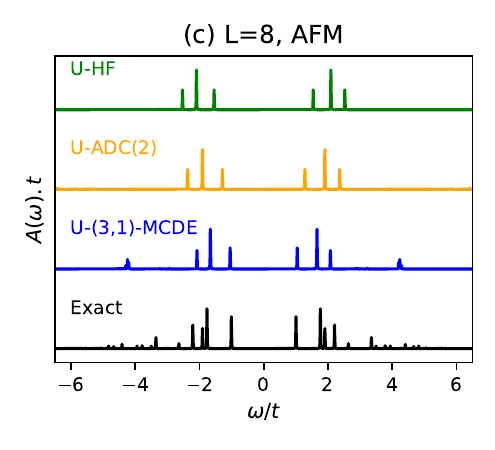}
    \vspace{0.02\linewidth}
    \includegraphics[width=0.47\linewidth]{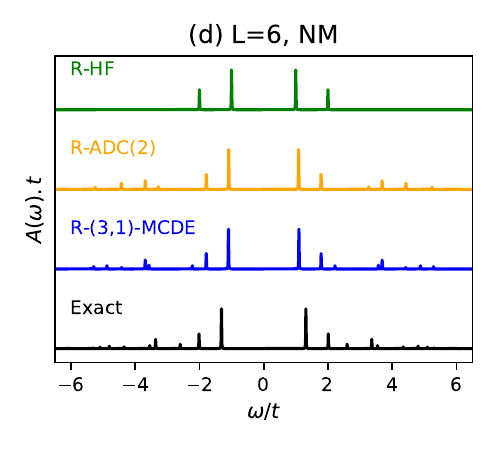}
    \caption{Spectral function obtained for $U/t = 4$ with $L = N = 4$ (a), $L = N = 6$ (b), 
and $L = N = 8$ (c), using the unrestricted HF reference state. Panel (d) shows, 
for comparison, the result obtained for $L = N = 6$ within the symmetry-restricted 
approach, i.e. imposing $S = 0$ at the HF level. All spectral functions are 
normalized to their maximum value.}
    \label{fig:6_un}
\end{figure*}

Working at half-filling, the number of sites $L$ equals the number of particles $N$ by 
definition. The eigenvectors of the HF one-body Hamiltonian associated with the $N$ lowest 
eigenvalues define the occupied (hole) states whereas the remaining $N$ 
eigenvectors define the unoccupied {(electron)} states. From these, {$2e1h$} and {$1e2h$} many-body configurations are constructed. There are $2N\binom{N}{2}$ such configurations, and the total 
dimensionality of the ISC space is therefore $2N\left(1+\binom{N}{2}
\right)$.

With the HF one-body basis at hand, coupling and interaction matrices given in Eqs.~\eqref{eq:selfen3} and {~\eqref{connectionmat}} are computed in order to build and diagonalize the (3,1)-MCDE and ADC(2) effective Hamiltonians. Retrieving the approximate Green's  function $g(\omega)$ through Eq.~\eqref{eq:MCDEGREEN}, the spectral function is subsequently obtained as 
\begin{equation}
    A(\omega) \equiv \frac{1}{\pi} \textbf{Tr} \left| \text{Im}\, 
    g(\omega) \right|\, .
\end{equation}

The U-HF, U-ADC(2) and U-(3,1)-MCDE spectral functions are compared to exact results in Fig.~\ref{fig:6_un} for $L\in\{4,6,8\}$. For $L=6$, panel (d) also displays results based on the R-HF reference state. The {spectral} functions being symmetric with respect to  $\omega = 0$, due to the  even number of particles and half filling, the discussion can be restricted to $\omega \ge 0$.
    
The exact {spectral} functions typically present two or three dominant peaks supplemented by a set of smaller peaks, the "satellites". While the quasiparticle peaks are already present in the HF solution with or without symmetry breaking, the satellites are entirely absent{, as expected}.  Going beyond  HF  by adding {$2e1h$} and {$1e2h$} doorway many-body states leads to the appearance of {low-strength satellite peaks} at intermediate and high $\omega$ values. In particular, U-(3,1)-MCDE results display more fragmentation than U-ADC(2) ones. Still, even the  U-(3,1)-MCDE spectral function does not grasp the full complexity of the exact solution although it shows a better spectrum than ADC(2).

\subsection{R-HF versus U-HF based results for $L=6$}
\label{sec:break}

Focusing on $L=6$ in Fig.~\ref{fig:6_un}, it is interesting to observe that results based on the R-HF state display more fragmentation than when using the U-HF reference state, leading to a better agreement with the exact solution. Let us now analyze in more detail the quality of the results based on R-HF versus U-HF reference states.

\subsubsection{Quasiparticle peaks}

\begin{figure*}
    \centering
    \includegraphics[width=0.9\linewidth]{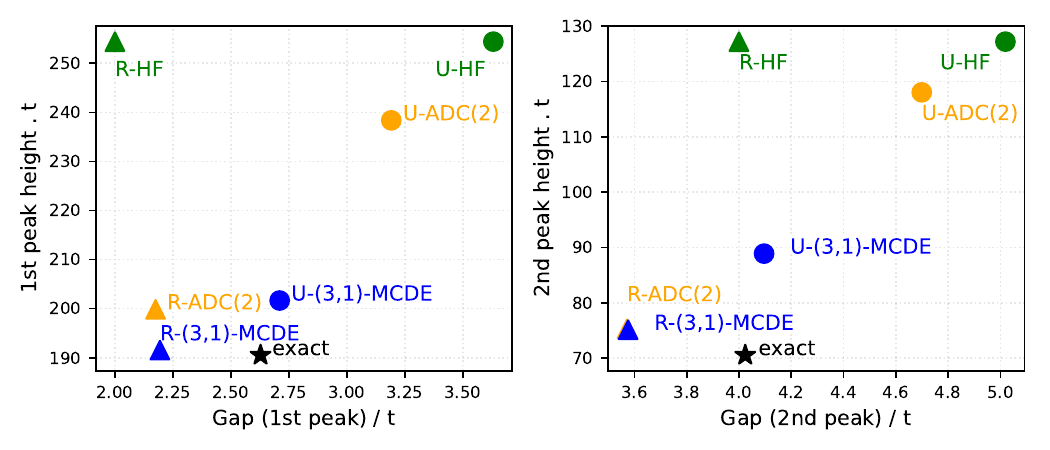}
    \caption{Gap-height plane for the first (left) and second (right) peaks at $U/t=4$ for $L=6$ deduced from Fig. \ref{fig:6_un}. Each point corresponds to a given 
    approximation scheme shown in the different panels of this figure, 
    while the exact result is indicated by a star. 
    The distance to the exact point serves as a measure of 
    accuracy.}
    \label{fig:gapheight}
\end{figure*}

Since they constitute the dominant contributions to the spectral function, let us focus first on the properties of the two quasiparticle peaks. Defining the energy gap of a given peak as twice the $\omega$ value of the peak, Fig.~\ref{fig:gapheight} reports it for the first and second quasiparticle peaks as a function of the height of the peaks for the different calculations of interest.

Both R-HF and U-HF lead to quasiparticle peaks whose heights are significantly overestimated. As for the energy gaps, U-HF overestimates them whereas R-HF underestimates them, although to a lesser extent; for instance, the position of the second peak is perfectly reproduced by R-HF.

Going beyond the HF approximation typically improves the agreement with exact results by decreasing the height of the peaks in all cases{, due to the appearance of satellites,} and by decreasing (increasing) the energy gap when starting from the U-HF (R-HF) reference state. While the improvement remains modest for U-ADC(2), U-(3,1)-MCDE {offers} a significant impact, delivering the best energy gaps of all tested truncation schemes. Starting from the R-HF reference state, ADC(2) and (3,1)-MCDE deliver similar corrections, leading to an excellent reproduction of both peak heights as well as a decent estimation, although underestimated, of the energy gaps.

Interestingly, while  R-ADC(2) and U-ADC(2) deliver very different results, R-(3,1)-MCDE and U-(3,1)-MCDE provide more consistent predictions, i.e. the superiority of (3,1)-MCDE over ADC(2) can be judged both by the better reproduction of exact results it offers but also in view of its stronger robustness associated with a larger insensitivity to the employed reference state.
\begin{figure*}
    \centering
    \includegraphics[width=0.9\linewidth]{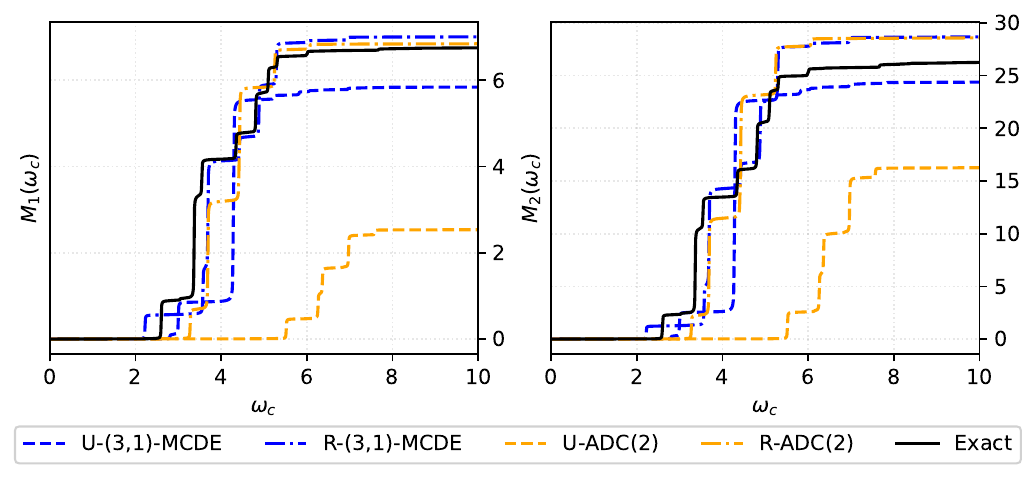}
    \caption{Cumulative spectral moments $M_1(\omega_c)$ (left) and $M_2(\omega_c)$ (right) 
%of the incoherent part of the spectral function 
as a function of the 
cutoff frequency $\omega_c$, for the 6-site Hubbard ring at $U/t = 4$. Note that, the HF results (not shown here), essentially lead to very small moments. 
%The moments are computed from the positive satellites only, i.e. excluding the main collective peaks seen in Fig. \ref{fig:6_un}. 
}
    \label{fig:moments}
\end{figure*}
\begin{comment}
    
This figure clearly illustrated that the (3,1)-MCDE seems to systematically outperforms ADC(2): its point always lies 
closer to the exact point in the gap-height plane
%, indicating a consistent improvement of the approximation. 
%It is shown in Fig.~\ref{fig:gapheight} for $U/t=4$. 
The choice of reference state introduces a trade-off between the 
accuracy of the gap and that of the peak heights. On the one hand, 
the AFM reference state yields more accurate gaps, as reflected by 
the smaller horizontal distance between the exact point and U-(3,1)-MCDE 
compared to R-(3,1)-MCDE. On the other hand, the NM reference state better 
reproduces the peak heights, as reflected by the smaller vertical 
distance between the exact point and R-(3,1)-MCDE compared to U-(3,1)-MCDE. 
\end{comment}

\subsubsection{Satellites}

To best quantify the capacity of a given approximation to reproduce the key characteristics of the subleading satellites, the cumulative moments of the {\it incoherent} part $A_{\text{inc}}(\omega)$ of the spectral function, i.e. only including contributions from the positive satellites, are introduced. These moments are defined as a cumulative weighted sum of $A_{\text{inc}}(\omega)$
\begin{equation}
%    \mathbf{M}_k = \int_\mathbb{R}d\omega\,A(\omega)\omega^k\,.
    \mathbf{M}_k(\omega_c) = \int_0^{\omega_c} d\omega\,\omega^k \, A_{\text{inc}}(\omega)\, ,
\end{equation}
where $\omega_c$ denotes the energy up to which the running sum is computed, which can be varied from zero to infinity, and the natural number $k$ denotes the order of the moment. Moments present a sudden jump up at the location of each (group of) satellite, followed by a plateau. Moments eventually saturate to a maximal value at an energy denoted as $\Omega_c$. By construction, the total moment value is characteristic of the amount of strength carried by the satellite contributions, knowing that augmenting $k$ increasingly emphasizes high-energy satellite {(i.e. non-collective)} contributions. More specifically, the normalized first moment $M_1/M_0$ provides a measure of the average energy of the satellite distribution, while the normalized second moment $M_2/M_0$ contains additional information on its energy spreading. Indeed, the variance of the distribution can be obtained as $\sigma^2 = M_2/M_0-(M_1/M_0)^2$. Therefore, comparing the first and second moments allows one to separately assess the ability of a given approximation to reproduce the average energy localization and the energy spreading of the satellite strength.

The first ($k=1$) and second ($k=2$) moments are compared in Fig.~\ref{fig:moments} to exact results for the different approximations shown in panels (b) and (d) of Fig. \ref{fig:6_un}. The relative errors on the full moments are reported in {Table}~\ref{tab:relmom}.

First, the moments pinpoint that the U-ADC(2) is the least performing approach to describe the energy distribution and strength of the satellites. Going to U-(3,1)-MCDE, the full moments are largely improved even though the fragmentation of the strength is strongly underestimated such that the energy localization of individual peaks is not well reproduced. 

Calculations based on the R-HF reference state clearly outperform those based on the U-HF one and lead to a very good reproduction of the progressive rise of cumulative moments as a function of $\omega_c$. Eventually, the total $M_1$ value is perfectly reproduced by both  R-ADC(2) and  R-(3,1)-MCDE. Looking in detail, the localization of individual satellites is better reproduced by the R-(3,1)-MCDE calculation, especially for $\omega_c > 3$, i.e. neglecting interaction matrices in ADC(2) leads to an underestimated fragmentation, without changing the asymptotic cumulative strength.

\begin{table}[t]
\begin{ruledtabular}
\begin{tabular}{lcc}
  & $M_1$ (\%) & $M_2$ (\%)  \\
\hline
 U-ADC(2)  & 62.42 & 38.04  \\
 U-(3,1)-MCDE    & 13.48 & 7.16   \\
 R-ADC(2)   & 1.38  & 8.82   \\
 R-(3,1)-MCDE     & 3.76  & 9.19   \\
\end{tabular}
\end{ruledtabular}
\caption{Relative errors (in \%) with respect to the exact values of the maximal value of $M_1$ and $M_2$ shown in Fig. \ref{fig:moments}, for different methods and reference states with $L=N=6$ and $U/t=4$.}\label{tab:relmom}
\end{table}

\subsection{Performance as a function of $U/t$}

In order to provide a wider perspective on the performance of the approximation schemes under consideration, the discussion is now extended to a large interval of coupling strengths. Particular attention will be paid to the accuracy of the gap, which is the central observable for Mott physics. Furthermore, two additional metrics of relative errors are introduced. The first, $D_{\rm Main}^i$, measures the average relative error on the gap and height of the $i$-th quasiparticle peak, while the second, $D_{\rm Sat}$, measures the average relative error on the first- and second-order moments of the satellite peaks 
\begin{subequations}
\label{errormetrics}
\begin{eqnarray}
D_{\rm Main}^i &\equiv& \frac{1}{2} \left( \frac{|G^i - \tilde{G}^i|}{G^i} +
\frac{|H^i - \tilde{H}^i|}{H^i} \right), \label{eq:dmain} \\
D_{\rm Sat} &\equiv& \frac{1}{2} \left( \frac{|M_1 - \tilde{M}_1|}{M_1} +
\frac{|M_2 - \tilde{M}_2|}{M_2} \right). \label{eq:sat}
\end{eqnarray}
\end{subequations}
Here, $G^i$ and $H^i$ denote the exact gap and exact height of the $i$-th quasiparticle peak, while $M_1$ and $M_2$ denote the exact first- and second-order moments, respectively. The quantities $\tilde{G}^i$, $\tilde{H}^i$, $\tilde{M}_1$, and $\tilde{M}_2$ represent their corresponding approximations. These metrics provide a  meaningful criterion to assess the systematic behavior across all values of $U/t$. 
The gap errors are reported in Fig.~\ref{fig:errorgap6} for $L=6$ and Fig.~\ref{fig:errorgap4} for $L=4,8$. The metrics are displayed in Fig.~\ref{fig:error_4} for the 4-, 6- and 8-sites Hubbard ring.

\subsubsection{Closed-shell system ($L=6$)}

\begin{figure}[t]
    \centering
    \includegraphics[width=0.9\linewidth]{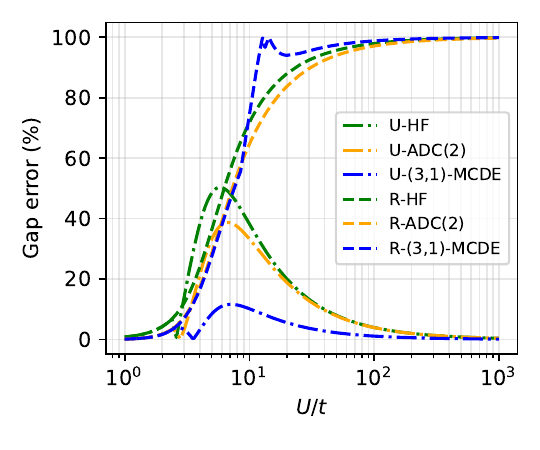}
    \caption{{Relative gap error as a function of $U/t$ for the different approximations discussed in this work. Results are shown for $L=N=6$ in percentage of error compared to the exact case.}}
    \label{fig:errorgap6}
\end{figure}

Focusing on the gap, Fig.~\ref{fig:errorgap6} illustrates the accuracy of the approaches over a wide range of $U/t$ for $L=6$. U-(3,1)-MCDE provides the best description, except in a narrow window directly after the phase transition, where U-ADC(2) appears more accurate. These unrestricted approaches are exact in the band limit ($U/t \rightarrow 0$) and atomic limit ($U/t \rightarrow \infty$). In addition, the unrestricted error nearly vanishes at an intermediate point. This point is a crossing at which U-ADC(2) and U-(3,1)-MCDE switch from underestimating to overestimating the gap; its location carries no particular physical meaning. Finally, since the restricted reference is independent of $U/t$, the figure also highlights that the restricted approximation increasingly fails to reproduce the gap as $U/t$ grows.

\begin{figure*}
    \centering
    \includegraphics[width=\linewidth]{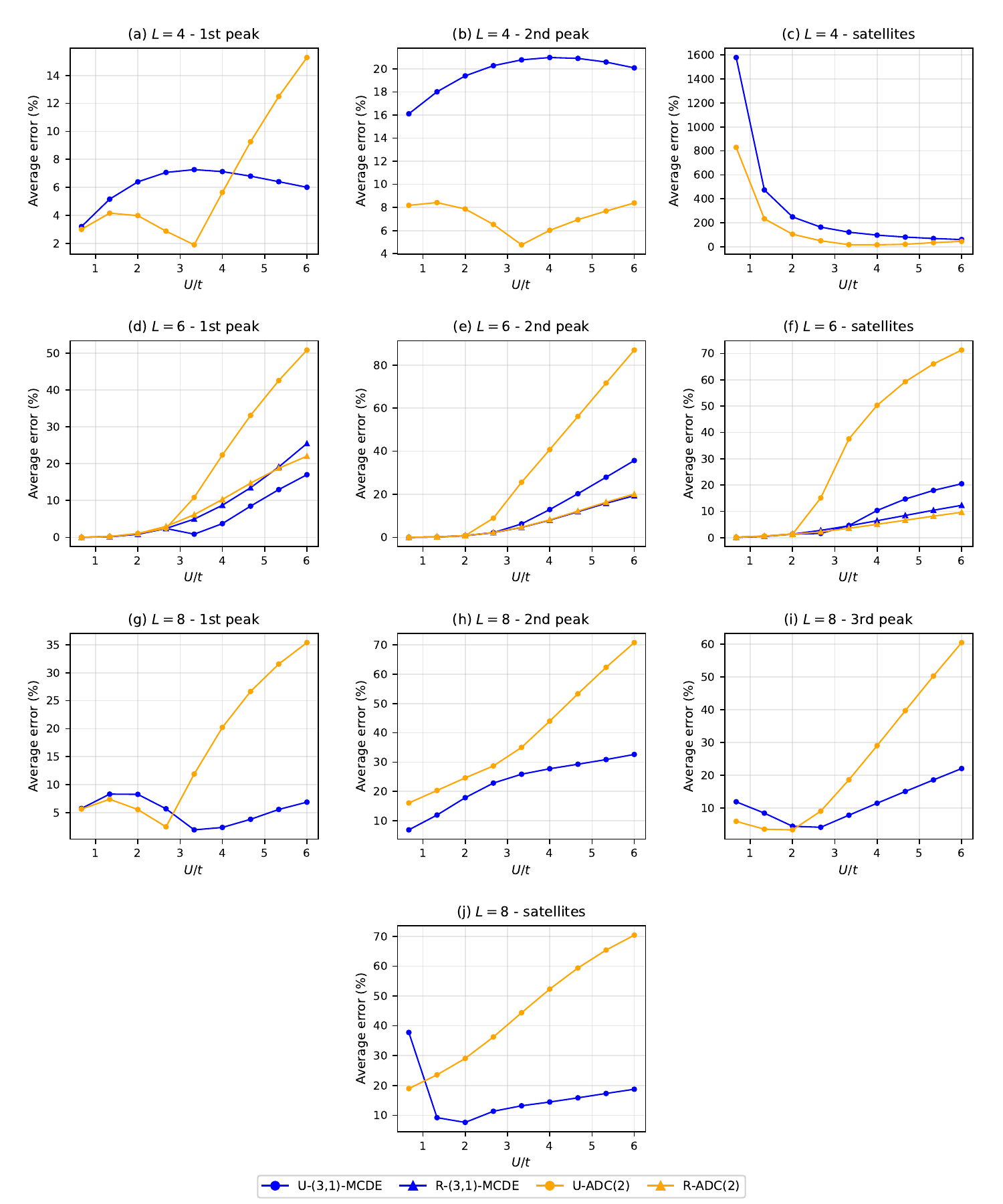}
    \caption{Relative error metrics {$D^1_{\rm Main}$ ((a), (d), (g)), $D^2_{\rm Main}$ ((b), (e), (h)), $D^3_{\rm Main}$ ((i)) and  $D_{\rm Sat}$ ((c), (f), (j)), defined in Eqs.~\eqref{errormetrics}} as function of $U/t$ for the different approximations discussed in this work. Results are shown for $L=N\in \{4,6,8\}$ in percentage of error compared to the exact case.}
    \label{fig:error_4}
\end{figure*}

The error metrics defined in Eq.~\eqref{errormetrics} are displayed in panel (d) to (f) of Fig.~\ref{fig:error_4} as a function of $U/t$ for $L=6$. 
All approximation schemes perform very well below $U/t \sim 2$ and lead to similar results. This is expected given that U-HF reduces R-HF below the phase transition and the problem becomes essentially perturbative in the weak coupling regime. 
For $U/t > 2$, the error is increasing monotonically for all approximation schemes, though not at the same rate.  
The U-ADC(2) approximation always delivers the largest error for both the quasiparticle peaks and the satellites. Going to either the R-HF reference state or to the (3,1)-MCDE truncation improve the situation significantly and systematically. Regarding the first quasiparticle peak, U-(3,1)-MCDE delivers the best answer for all interaction strengths, R-ADC(2) and R-(3,1)-MCDE being slightly less performing and very close to one another. The situation is reversed for the second quasiparticle peak and the satellites, i.e. using the R-HF reference state constitutes the best option.

\subsubsection{Open-shell systems ($L=4$ and $8$)}
%\subsection{8-sites Hubbard ring}
\label{sec:perf}

% \begin{figure*}
%     \centering
%     %\includegraphics[width=\linewidth]{images/errors_all_horizontal_linear4.pdf}
%     \includegraphics[width=0.9\linewidth]{images/error_8_all_qphf31.pdf}% or error_8_all_qphf.pdf 
%     \caption{\textit{Same as Fig. \ref{fig:error_sys} for the open-shell $N=L=8$ case. In this case, since there are three quasiparticle peaks in panel (c) of Fig. \ref{fig:6_un}, we added the error on the third peak.  }}
%     \label{fig:error_8}
% \end{figure*}

Focusing first on the $L=8$ case (panels (g) to (j) of Fig.~\ref{fig:error_4}), U-(3,1)-MCDE is seen to outperform U-ADC(2), except at very low $U/t$ values. A clear difference with the closed-shell case is the non-negligible 
error observed even at very small coupling. This is due to the fact that the perturbative behavior at small $U/t$ is missed by the approximate calculations due to the U-HF reference state leading to an exaggerated fragmentation compared to the exact result that exhibits very little of it. This leads to an overestimation of $D_{\mathrm{Sat}}$. As $U/t$ increases, the exact solution itself develops significant fragmentation, giving rise to a crossover where U-ADC(2) and U-(3,1)-MCDE deliver a realistic fragmentation before underestimating it at even higher $U/t$ values.

As the coupling increases, the errors indeed rise again, with a transition occurring around $U/t \sim 2$--$3$. Above this transition, the behaviors are similar to what was observed in the closed-shell case. The order of magnitude of the errors when $U/t$ approaches $6$ are similar also to the latter case. Again, in this regime, U-(3,1)-MCDE captures better both the quasiparticle peaks and satellite behaviors than U-ADC(2). The most striking feature of Fig.~\ref{fig:error_4} is the strong divergence of \mbox{U-ADC(2)} 
at large $U/t$. This suggests that the absence of the interaction matrices $\Sigma^{2e1h}$ and $\Sigma^{1e2h}$ in ADC(2)  becomes increasingly critical as $U/t$ becomes large, { as one would expect.}

Let us finally consider the $L=4$. As visible from panels (a) to (c) of Fig.~\ref{fig:error_4}, the non-trivial evolution as a function of $U/t$ already observed for $L=8$ is further amplified in this smaller lattice. For the first peak, U-ADC(2) again exhibits a lower error than U-(3,1)-MCDE all the way to $U/t \sim 4$ where a crossing occurs. For the second peak and the satellite moments, U-ADC(2) systematically outperforms U-(3,1)-MCDE over the full interaction range, although the associated errors remain very large. As in the $L=8$ case, this is attributed to the U-HF reference state that generates an incorrect fragmentation of the spectral function. A crossover between weak and strong coupling regimes is expected but  not clearly observed. This can be understood from the fact that, in the $L=4$ case, the exact spectral function remains only weakly fragmented even at large $U/t$. Consequently, symmetry-unrestricted approaches systematically overestimate spectral fragmentation even at strong coupling.
 
 \begin{figure}[t]
    \centering
    \includegraphics[width=0.9\linewidth]{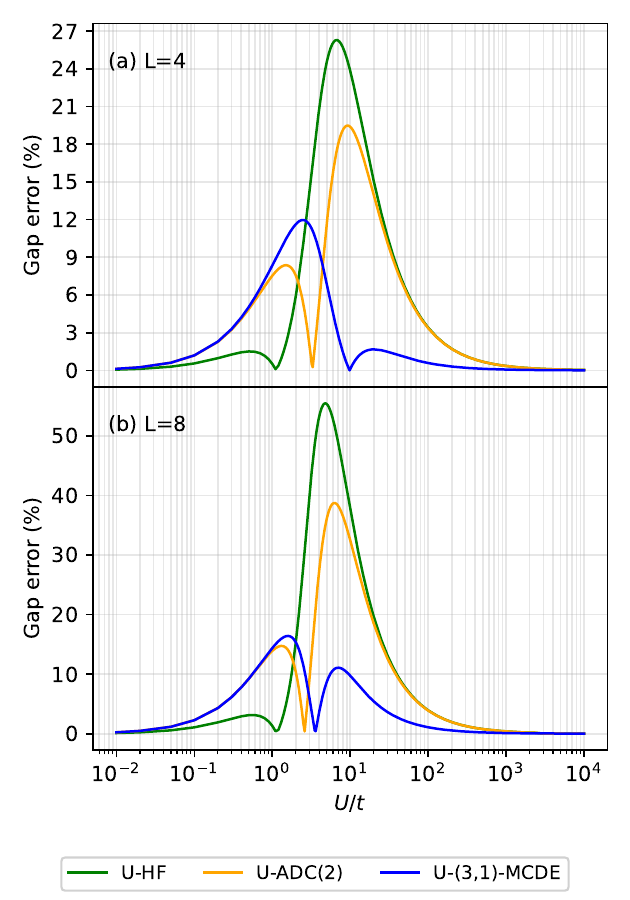}
    \caption{Relative gap error as a function of $U/t$ for the different approximations discussed in this work. {Results are shown for $L=N=4$ ((a)) and  $L=N=8$ ((b))} in percentage of error compared to the exact case.}
    \label{fig:errorgap4}
\end{figure}
 Focusing on the energy gap, Fig.~\ref{fig:errorgap4} shows that the same conclusion as for the 6-site is reached. The gap is reproduced almost exactly in the band limit $U/t \rightarrow 0$, the atomic limit $U/t \rightarrow \infty$, and at an intermediate point for the 4- and 8-sites ring. Besides, U-ADC(2) performs better than U-(3,1)-MCDE for $U/t$ smaller than $3.5$ when $L=4$ and $2$ when $L=8$.

Finally, these results highlight that both U-ADC(2) and U-(3,1)-MCDE approximation schemes exhibit significant limitations for open-shell systems in the weak-coupling regime or for very small system sizes. Beyond these particular limitations, U-(3,1)-MCDE constitutes a well-performing approximation to describe both dominant and satellite structures that systematically outperforms U-ADC(2).

\section{Conclusions}
\label{conclusions}

In this work we establish a formal equivalence between Green's function calculations based on the so-called (3,1)-MCDE approximation on the one hand and an extension of the ADC(2) approximation, similar to ADC(2)-X, on the other hand. This  result provides a rigorous bridge between two independently developed frameworks to design efficient approximations to the one-body self energy in fermionic systems. In particular, it {further} clarifies the diagrammatic content of the recently proposed (3,1)-MCDE approximation that happens to lie in between the well-known ADC(2) and ADC(3) truncations.

This equivalence has an immediate practical consequence: it enables the 
extension of the (3,1)-MCDE to open-shell systems by leveraging the 
existing ADC machinery. To illustrate this, both ADC(2) 
and (3,1)-MCDE were benchmarked on 4-, 6- and 8-site Hubbard rings at half-filling, where the HF ground state exhibits {a} closed-shell structure for $L = 6$ and an open-shell structure for $L=4,8$. The difficulty in dealing with open-shell systems is circumvented by using a spin-unrestricted HF reference state, which introduces a mean-field NM-AFM transition and yields a nondegenerate AFM reference state amenable to standard Green's function calculations on top of it.

The numerical benchmarks reveal a consistent picture. For the quasi-particle 
peaks, (3,1)-MCDE systematically outperforms ADC(2), with the improvement 
becoming more robust as the system size increases from $L = 6$ to $L = 8$, 
suggesting that (3,1)-MCDE becomes increasingly accurate in the thermodynamic 
limit. This trend is particularly striking in the strong-coupling regime, 
where the absence of the interaction matrices $\Sigma^{2e1h}$ and $\Sigma^{1e2h}$ in ADC(2) 
leads to a systematic divergence of the error metrics, while (3,1)-MCDE 
remains well-behaved. For the satellites, the results are more 
nuanced: (3,1)-MCDE improves upon ADC(2) for the spectral moments in most 
cases, but the choice of reference state introduces a trade-off between 
the accuracy of the average satellite energy and that of the satellite 
fluctuations.

Several directions for future work naturally follow from these results. 
First, a dedicated analysis of the $L = 4$ case, whose anomalous behavior 
suggests that additional physical mechanisms are at play, would provide 
valuable insight into the limitations of the unrestricted approach. 
Second, the extension of the present benchmarks to larger system sizes 
would allow a more systematic assessment of the thermodynamic limit 
behavior. Third, the application of (3,1)-MCDE to realistic open-shell 
molecular systems and open-shell nuclei, now made possible by the ADC(2)-X 
equivalence, constitutes a natural next step. Fourth, the framework 
introduced in Appendix~B opens a route toward higher-order (n,s)-MCDE
approximations, by systematically including higher-order coupling and 
interaction matrices. Finally, a better understanding of the interplay 
between symmetry breaking and spectral accuracy --- in particular the 
origin of the trade-off between gap and peak-height accuracy --- would 
provide valuable guidance for the choice of optimized reference states in future 
applications.

\begin{acknowledgments}
PR and JAB thank the French Agence Nationale de la Recherche (ANR) for financial support (Grant Agreement ANR-22-CE30-0027).
\end{acknowledgments}

\bibliography{aipsamp}% Produces the bibliography via BibTeX.

\newpage
\appendix

\begin{widetext}

\section{ADC building blocks: matching to perturbation theory}
\label{app:adc_matchings}

The compact ADC expressions used throughout the main text are presently mapped
onto the perturbative expansion of the dynamical self-energy written in terms of quasiparticle amplitudes $\mathcal{X},\mathcal{Y}$ and antisymmetrized interaction matrix elements of $\bar v$. This appendix provides explicit formulas for the ADC(2/3) coupling and interaction matrices. 

\subsection{ADC(2): identification of $\mathcal{M}^{(1)}$ and $\mathcal{N}^{(1)}$}

At second order, the ADC self-energy in intermediate-state representation reads
\begin{equation}
\label{eq:order2_app}
\tilde\Sigma^{(2)}_{\alpha\beta}(\omega)
=
\sum_{r}\mathcal{M}^{(1)\dagger}_{\alpha r}
\left[\frac{1}{\omega-E_r^>+i\eta}\right]
\mathcal{M}_{r \beta}^{(1)}
+
\sum_{s}\mathcal{N}^{(1)}_{\alpha s}
\left[\frac{1}{\omega-E_s^<-i\eta}\right]
\mathcal{N}_{s \beta}^{(1)\dagger}\, ,
\end{equation}
where $r$ and $s$ label $2e1h$ and $1e2h$ intermediate states, respectively.
This expression is matched onto the standard second-order perturbative self-energy~\cite{Barbieri:2016uib}
\begin{equation}
\label{eq:ADC2}
\begin{aligned}
\tilde\Sigma^{(2)}_{\alpha\beta}(\omega)
=& \frac{1}{2}\sum_{\epsilon\gamma\delta}\sum_{\lambda\mu\rho}\bar{v}_{\alpha\epsilon\gamma\delta}\Bigg(
\sum_{n_1n_2k_3}\frac{(\mathcal{X}_\gamma^{n_1}\mathcal{X}_\delta^{n_2}\mathcal{Y}_\epsilon^{k_3})^*
\,\mathcal{X}^{n_1}_\lambda\mathcal{X}^{n_2}_\mu\mathcal{Y}_\rho^{k_3}}
{\omega-(\varepsilon^+_{n_1}+\varepsilon^+_{n_2}-\varepsilon^-_{k_3})+i\eta}\\
&\hspace{3.0em}
+\sum_{k_1k_2n_3}\frac{\mathcal{Y}_\gamma^{k_1}\mathcal{Y}_\delta^{k_2}\mathcal{X}_\epsilon^{n_3}\,
(\mathcal{Y}^{k_1}_\lambda\mathcal{Y}^{k_2}_\mu\mathcal{X}_\rho^{n_3})^*}
{\omega-(\varepsilon^-_{k_1}+\varepsilon^-_{k_2}-\varepsilon^+_{n_3})-i\eta}
\Bigg)\bar{v}_{\lambda\mu\beta\rho}\, .
\end{aligned}
\end{equation}
The prefactor $1/2$ disappears if the sums are restricted to ordered pairs, e.g.\ $n_1<n_2$ and
$k_1<k_2$, which is equivalent to using antisymmetrized $2e$ and $2h$ states in the definition of
the composite indices $r=(n_1n_2k_3)$ and $s=(k_1k_2n_3)$.
With this convention, the matching identifies the first-order coupling matrices as
\begin{subequations}
\label{eq:couplings_order1}
\begin{equation}
\label{eq:M1_def}
\mathcal{M}^{(1)}_{r\beta}
\equiv
\mathcal{M}^{(1)}_{(n_1n_2k_3)\beta}
=
-\sum_{\lambda\mu\rho}\bar{v}_{\lambda\mu\beta\rho}\,
\mathcal{X}^{n_1}_\lambda\mathcal{X}^{n_2}_\mu\mathcal{Y}_\rho^{k_3}\, ,
\end{equation}
\begin{equation}
\label{eq:N1_def}
\mathcal{N}^{(1)\dagger}_{s\beta}
\equiv
\mathcal{N}^{(1)\dagger}_{(k_1k_2n_3)\beta}
=
-\sum_{\lambda\mu\rho}\bar{v}_{\lambda\mu\beta\rho}\,
\big(\mathcal{Y}^{k_1}_\lambda\mathcal{Y}^{k_2}_\mu\mathcal{X}_\rho^{n_3}\big)^*\, .
\end{equation}
\end{subequations}

\subsection{ADC(3): identification of $\mathcal{C}^{(1)}$, $\mathcal{D}^{(1)}$, $\mathcal{M}^{(2)}$ and $\mathcal{N}^{(2)}$}

At third order, the expanded ADC self-energy can be organized into two distinct contributions:
(i) cross terms involving the second-order couplings $\mathcal{M}^{(2)}$, $\mathcal{N}^{(2)}$ with
the zeroth-order propagators in the intermediate spaces, and (ii) terms where the propagation
within the $2e1h$ and $1e2h$ manifolds is dressed by the first-order interaction matrices
$\mathcal{C}^{(1)}$ and $\mathcal{D}^{(1)}$. Explicitly,
\begin{equation}
\label{eq:order3_app}
\begin{aligned}
\tilde\Sigma^{(3)}_{\alpha\beta}(\omega)
=& \sum_{r}\mathcal{M}^{(2)\dagger}_{\alpha r}
\left[\frac{1}{\omega-E_r^>+i\eta}\right]\mathcal{M}_{r \beta}^{(1)}
+\sum_{r}\mathcal{M}^{(1)\dagger}_{\alpha r}
\left[\frac{1}{\omega-E_r^>+i\eta}\right]\mathcal{M}_{r \beta}^{(2)}\\
&+\sum_{rr'}\mathcal{M}^{(1)\dagger}_{\alpha r}
\left[\frac{1}{\omega-E_r^>+i\eta}\right]\mathcal{C}^{(1)}_{rr'}
\left[\frac{1}{\omega-E_{r'}^>+i\eta}\right]\mathcal{M}_{r' \beta}^{(1)}\\
&+\sum_{s}\mathcal{N}^{(2)}_{\alpha s}
\left[\frac{1}{\omega-E_s^<-i\eta}\right]\mathcal{N}_{s \beta}^{(1)\dagger}
+\sum_{s}\mathcal{N}^{(1)}_{\alpha s}
\left[\frac{1}{\omega-E_s^<-i\eta}\right]\mathcal{N}_{s \beta}^{(2)\dagger}\\
&+\sum_{ss'}\mathcal{N}^{(1)}_{\alpha s}
\left[\frac{1}{\omega-E_s^<-i\eta}\right]\mathcal{D}^{(1)}_{ss'}
\left[\frac{1}{\omega-E_{s'}^<-i\eta}\right]\mathcal{N}_{s' \beta}^{(1)\dagger}\, .
\end{aligned}
\end{equation}
Here $r$ and $r'$ label $2e1h$ configurations and $s$ and $s'$ label $1e2h$ configurations, with the
same ordering conventions as in the second-order case.

Matching Eq.~\eqref{eq:order3_app} to the third-order perturbative self-energy yields the
first-order interaction matrices within the intermediate-state spaces~\cite{Barbieri:2016uib},
\begin{subequations}
\label{eq:CD1_defs}
\begin{equation}
\label{eq:C1_def}
\begin{aligned}
\mathcal{C}^{(1)}_{rr'}
\equiv
\mathcal{C}^{(1)}_{(n_1n_2k_3) (n_1'n_2'k_3')}
=& -\sum_{\alpha\beta\gamma\delta}\Big[
\mathcal{X}_\alpha^{n_1}\mathcal{X}_\beta^{n_2}\,
\bar{v}_{\alpha\beta\gamma\delta}\,
\big(\mathcal{X}_\gamma^{n_1'}\mathcal{X}_\delta^{n_2'}\big)^*\,
\delta_{k_3 k_3'} \\
&\quad
+\mathcal{X}_\alpha^{n_1}\mathcal{Y}_\beta^{k_3}\,
\bar{v}_{\alpha\delta\beta\gamma}\,
\big(\mathcal{X}_\gamma^{n_1'}\mathcal{Y}_\delta^{k_3'}\big)^*\,
\delta_{n_2 n_2'} \\
&\quad
-\mathcal{X}_\alpha^{n_2}\mathcal{Y}_\beta^{k_3}\,
\bar{v}_{\alpha\delta\beta\gamma}\,
\big(\mathcal{X}_\gamma^{n_1'}\mathcal{Y}_\delta^{k_3'}\big)^*\,
\delta_{n_1 n_2'} \\
&\quad
-\mathcal{X}_\alpha^{n_1}\mathcal{Y}_\beta^{k_3}\,
\bar{v}_{\alpha\delta\beta\gamma}\,
\big(\mathcal{X}_\gamma^{n_2'}\mathcal{Y}_\delta^{k_3'}\big)^*\,
\delta_{n_2 n_1'} \\
&\quad
+\mathcal{X}_\alpha^{n_2}\mathcal{Y}_\beta^{k_3}\,
\bar{v}_{\alpha\delta\beta\gamma}\,
\big(\mathcal{X}_\gamma^{n_2'}\mathcal{Y}_\delta^{k_3'}\big)^*\,
\delta_{n_1 n_1'}
\Big]\, ,
\end{aligned}
\end{equation}
\begin{equation}
\label{eq:D1_def}
\begin{aligned}
\mathcal{D}^{(1)}_{ss'}
\equiv
\mathcal{D}^{(1)}_{(k_1k_2n_3) (k_1'k_2'n_3')}
=& -\sum_{\alpha\beta\gamma\delta}\Big[
-\big(\mathcal{Y}_\alpha^{k_1} \mathcal{Y}_\beta^{k_2}\big)^*\,
\bar{v}_{\alpha\beta\gamma\delta}\,
\mathcal{Y}_\gamma^{k_1'} \mathcal{Y}_\delta^{k_2'}\,
\delta_{n_3 n_3'} \\
&\quad
-\big(\mathcal{Y}_\alpha^{k_1} \mathcal{X}_\beta^{n_3}\big)^*\,
\bar{v}_{\alpha\delta\beta\gamma}\,
\mathcal{Y}_\gamma^{k_1'} \mathcal{X}_\delta^{n_3'}\,
\delta_{k_2 k_2'} \\
&\quad
+\big(\mathcal{Y}_\alpha^{k_2} \mathcal{X}_\beta^{n_3}\big)^*\,
\bar{v}_{\alpha\delta\beta\gamma}\,
\mathcal{Y}_\gamma^{k_1'} \mathcal{X}_\delta^{n_3'}\,
\delta_{k_1 k_2'} \\
&\quad
+\big(\mathcal{Y}_\alpha^{k_1} \mathcal{X}_\beta^{n_3}\big)^*\,
\bar{v}_{\alpha\delta\beta\gamma}\,
\mathcal{Y}_\gamma^{k_2'} \mathcal{X}_\delta^{n_3'}\,
\delta_{k_2 k_1'} \\
&\quad
-\big(\mathcal{Y}_\alpha^{k_2} \mathcal{X}_\beta^{n_3}\big)^*\,
\bar{v}_{\alpha\delta\beta\gamma}\,
\mathcal{Y}_\gamma^{k_2'} \mathcal{X}_\delta^{n_3'}\,
\delta_{k_1 k_1'}
\Big]\, .
\end{aligned}
\end{equation}
\end{subequations}

In addition, the same matching fixes the second-order corrections to the coupling matrices
$\mathcal{M}^{(2)}$ and $\mathcal{N}^{(2)}$~\cite{Raimondi:2017mey},
\begin{subequations}
\label{eq:MN2_defs}
\begin{equation}
\label{eq:M2_def}
\begin{aligned}
\mathcal{M}^{(2)}_{r\alpha}
\equiv \mathcal{M}^{(2)}_{(n_1 n_2 k_3)\alpha}
=& \frac{1}{2}\sum_{k_4k_5}\sum_{\mu\nu\lambda}\sum_{\rho\sigma\gamma\delta}
\frac{\mathcal{X}_\rho^{n_1} \mathcal{X}_\sigma^{n_2}\,
\bar{v}_{\rho\sigma\gamma\delta}\,
\mathcal{Y}_\gamma^{k_4} \mathcal{Y}_\delta^{k_5}}
{\varepsilon_{k_4}^{-}+\varepsilon_{k_5}^{-}-\varepsilon_{n_1}^{+}-\varepsilon_{n_2}^{+}}\,
\big(\mathcal{Y}_\mu^{k_4} \mathcal{Y}_\nu^{k_5}\big)^* \mathcal{Y}_\lambda^{k_3}\,
\bar{v}_{\mu\nu\alpha\lambda}\\
&+\sum_{n_4k_5}\sum_{\mu\nu\lambda}\sum_{\rho\sigma\gamma\delta}
\frac{\mathcal{X}_\sigma^{n_2} \mathcal{Y}_\delta^{k_3}\,
\bar{v}_{\sigma\rho\delta\gamma}\,
\mathcal{Y}_\gamma^{k_5} \mathcal{X}_\rho^{n_4}}
{\varepsilon_{k_3}^{-}-\varepsilon_{n_2}^{+}+\varepsilon_{k_5}^{-}-\varepsilon_{n_4}^{+}}\,
\mathcal{X}_\mu^{n_1}\big(\mathcal{Y}_\nu^{k_5} \mathcal{X}_\lambda^{n_4}\big)^*\,
\bar{v}_{\mu\nu\alpha\lambda}\\
&-\sum_{n_4k_5}\sum_{\mu\nu\lambda}\sum_{\rho\sigma\gamma\delta}
\frac{\mathcal{X}_\sigma^{n_1} \mathcal{Y}_\delta^{k_3}\,
\bar{v}_{\sigma\rho\delta\gamma}\,
\mathcal{Y}_\gamma^{k_5} \mathcal{X}_\rho^{n_4}}
{\varepsilon_{k_3}^{-}-\varepsilon_{n_1}^{+}+\varepsilon_{k_5}^{-}-\varepsilon_{n_4}^{+}}\,
\mathcal{X}_\mu^{n_2}\big(\mathcal{Y}_\nu^{k_5} \mathcal{X}_\lambda^{n_4}\big)^*\,
\bar{v}_{\mu\nu\alpha\lambda}\, .
\end{aligned}
\end{equation}
\begin{equation}
\label{eq:N2_def}
\begin{aligned}
\mathcal{N}^{(2)}_{\alpha s}
\equiv \mathcal{N}^{(2)}_{\alpha(k_1 k_2 n_3)}
=& \frac{1}{2}\sum_{n_4n_5}\sum_{\mu\nu\lambda}\sum_{\rho\sigma\gamma\delta}
\bar{v}_{\alpha\lambda\mu\nu}\,
\mathcal{X}_\lambda^{n_3}\big(\mathcal{X}_\mu^{n_4} \mathcal{X}_\nu^{n_5}\big)^*
\frac{\mathcal{X}_\rho^{n_4} \mathcal{X}_\sigma^{n_5}\,
\bar{v}_{\rho\sigma\gamma\delta}\,
\mathcal{Y}_\gamma^{k_1} \mathcal{Y}_\delta^{k_2}}
{\varepsilon_{k_1}^{-}+\varepsilon_{k_2}^{-}-\varepsilon_{n_4}^{+}-\varepsilon_{n_5}^{+}}\\
&+\sum_{n_4k_5}\sum_{\mu\nu\lambda}\sum_{\rho\sigma\gamma\delta}
\bar{v}_{\alpha\lambda\mu\nu}\,
\big(\mathcal{Y}_\lambda^{k_5}\big)^* \mathcal{Y}_\mu^{k_1}\big(\mathcal{X}_\nu^{n_4}\big)^*\,
\frac{\mathcal{X}_\sigma^{n_4} \mathcal{Y}_\delta^{k_5}\,
\bar{v}_{\sigma\rho\delta\gamma}\,
\mathcal{Y}_\gamma^{k_2} \mathcal{X}_\rho^{n_3}}
{\varepsilon_{k_2}^{-}-\varepsilon_{n_3}^{+}+\varepsilon_{k_5}^{-}-\varepsilon_{n_4}^{+}} \\
&-\sum_{n_4k_5}\sum_{\mu\nu\lambda}\sum_{\rho\sigma\gamma\delta}
\bar{v}_{\alpha\lambda\mu\nu}\,
\big(\mathcal{Y}_\lambda^{k_5}\big)^* \mathcal{Y}_\mu^{k_2}\big(\mathcal{X}_\nu^{n_4}\big)^*\,
\frac{\mathcal{X}_\sigma^{n_4} \mathcal{Y}_\delta^{k_5}\,
\bar{v}_{\sigma\rho\delta\gamma}\,
\mathcal{Y}_\gamma^{k_1} \mathcal{X}_\rho^{n_3}}
{\varepsilon_{k_1}^{-}-\varepsilon_{n_3}^{+}+\varepsilon_{k_5}^{-}-\varepsilon_{n_4}^{+}}\, .
\end{aligned}
\end{equation}
\end{subequations}

\section{Spectral Representation Truncation}
\label{sec:adc-mcde}

This Appendix works out the relation between the (3,1)-MCDE Green's function and the corresponding approximate self-energy through the Dyson equation. It will be shown that the resulting approximation of the self-energy lies between ADC(2) and ADC(3) truncation orders. To do so, a new approximation scheme for the self-energy is introduced. Within this framework, it will be shown that the ADC(n) and (3,1)-MCDE approximations can be recovered as particular cases. 

The central idea of this approach is to truncate the spectral representation of the self-energy by limiting the order of the coupling and interaction matrices. More precisely, a {\it Spectral Representation Truncation} ($\mathrm{SRT}$) labeled by $(n,m,i)$ defines a self-energy in which the coupling matrices are retained up to order $n$, while the interaction matrices are retained up to order $m$. The third index, $i$, specifies the truncation level of the ISC space. That is, only the ISCs up to the $ie(i+1)h$ and $(i+1)eih$ sectors are retained.

As an illustration, the $\mathrm{SRT}(2,1,2)$ approximation corresponds to the self-energy
\begin{equation}
\label{eq:SRT(2,1,2)}
\begin{aligned}
\tilde\Sigma^{\mathrm{SRT}(2,1,2)}_{\alpha\beta}(\omega)
=&\sum_{rr'}
(\mathcal{M}^{(1)}+\mathcal{M}^{(2)})^\dagger_{\alpha r}
\left[\frac{1}{\omega-(E^>+\mathcal{C}^{(1)})+i\eta}\right]_{rr'}
(\mathcal{M}^{(1)}+\mathcal{M}^{(2)})_{r'\beta}\\
&+\sum_{ss'}
(\mathcal{N}^{(1)}+\mathcal{N}^{(2)})_{\alpha s}
\left[\frac{1}{\omega-(E^<+\mathcal{D}^{(1)})-i\eta}\right]_{ss'}
(\mathcal{N}^{(1)}+\mathcal{N}^{(2)})^\dagger_{s'\beta}\\
&+\sum_{qq'} \mathcal{M}^{(2)\dagger}_{\alpha q}
\left[\frac{1}{\omega-(E^{>}+\mathcal{C}^{(1)})+i\eta}\right]_{qq'}
\mathcal{M}^{(2)}_{q'\beta}
\\
&+
\sum_{pp'} \mathcal{N}^{(2)}_{\alpha p}
\left[\frac{1}{\omega-(E^{<}+\mathcal{D}^{(1)})-i\eta}\right]_{pp'}
\mathcal{N}^{(2)\dagger}_{p'\beta}\,,
\end{aligned}
\end{equation}
where, $q$ and $p$ are {\it composite} indices labeling second order ISCs, $2e3h$ and $3e2h$ configurations, respectively:
\begin{equation}
q \equiv (n_1 n_2 n_3 k_4 k_5)\, , \qquad p \equiv (k_1 k_2 k_3 n_4 n_5)\, .
\end{equation}

The truncation order of the ISC space is always less than or equal to that of the coupling matrix. This is because coupling an ISC of order $i$ to a single particle state requires at least $i$ interactions. Thus, to access $3e2h$ or $2e3h$ ISCs, a coupling matrix of order 2 is required. In the following, the shorthand notation $\mathrm{SRT}(n,m,1) = \mathrm{SRT}(n,m)$ will be used. It is straightforward to verify that the $\mathrm{SRT}(1,0)$ self-energy coincides with the ADC(2) self-energy given in Eq.~\eqref{eq:order2}, and that the $\mathrm{SRT}(2,1)$ self-energy corresponds to the ADC(3) self-energy given in Eq.~\eqref{eq:ADC3_Lehmann}. We will show below that the Green's function associated to the $\mathrm{SRT}(1,1)$ self-energy via Dyson's equation is a solution of the MCDE set.

\subsection{Resolvent}

The $\mathrm{SRT}(1,1)$ approximation of the self-energy is
\begin{equation}
\label{eq:MCDESE_11}
\begin{aligned}
\tilde\Sigma^{\text{SRT}(1,1)}_{\alpha\beta}(\omega)
=& \sum_{rr'}\mathcal{M}^{(1)\dagger}_{\alpha r}
\left[\frac{1}{\omega-(E^>+\mathcal{C}^{(1)})+i\eta}\right]_{rr'}
\mathcal{M}^{(1)}_{r' \beta}\\
&+\sum_{ss'}\mathcal{N}^{(1)}_{\alpha s}
\left[\frac{1}{\omega-(E^<+\mathcal{D}^{(1)})-i\eta}\right]_{ss'}
\mathcal{N}^{(1)\dagger}_{s'\beta}\, .
\end{aligned}
\end{equation}
\begin{comment}
As in Sec.~\ref{sec:mcde}, the forward ($2e1h$) and backward ($1e2h$) sectors are combined by introducing unified indices $u,v\in\{r,s\}$ and by defining
\begin{subequations}
\label{eq:O_G_B_defs}
\begin{align}
\label{eq:O_def}
\mathcal{O}^{(1)}_{u\beta}
\equiv&
\begin{cases}
\mathcal{M}^{(1)}_{r\beta} & \text{if } u\equiv r\,,\\[2pt]
\mathcal{N}^{(1)\dagger}_{s\beta} & \text{if } u\equiv s\,,
\end{cases}
\\[6pt]
\label{eq:Ginf_def}
G^{(\infty)}_{uv}(\omega)
\equiv&
\begin{cases}
\dfrac{\delta_{rr'}}{\omega-E_{r}^{>}+i\eta} & \text{if } (u,v)\equiv(r,r')\,,\\[10pt]
\dfrac{\delta_{ss'}}{\omega-E_{s}^{<}-i\eta} & \text{if } (u,v)\equiv(s,s')\,,\\[10pt]
0 & \text{otherwise}\,,
\end{cases}
\\[6pt]
\label{eq:B_def}
\mathcal{B}^{(1)}_{uv}
\equiv&
\begin{cases}
\mathcal{C}^{(1)}_{rr'} & \text{if } (u,v)\equiv(r,r')\,,\\[2pt]
\mathcal{D}^{(1)}_{ss'} & \text{if } (u,v)\equiv(s,s')\,,\\[2pt]
0 & \text{otherwise}\,.
\end{cases}
\end{align}
\end{subequations}
\end{comment}
In what follows, we adopt the Einstein summation convention: repeated indices are implicitly summed over. The resolvent can be used to write Eq.~\eqref{eq:MCDESE_11} as
\begin{equation}
\label{eq:MCDESE_schur_form}
\tilde\Sigma_{\alpha\beta}^{\mathrm{SRT}(1,1)}(\omega)
=
\mathcal{M}_{\alpha r}^{(1)\dagger}\,
\mathcal{R}_{rr'}^{(1)}(\omega)\,
\mathcal{M}_{r'\beta}^{(1)}+\mathcal{N}_{\alpha s}^{(1)}\,
\mathcal{R}_{ss'}^{(1)}(\omega)\,
\mathcal{N}_{s'\beta}^{(1)\dagger}\,,
\end{equation}
with the resolvent defined as
\begin{subequations}
\begin{equation}
\mathcal{R}^{(1)}_{rr'}(\omega)\equiv
\left[\left(G^{(\infty)}(\omega)\right)^{-1}-\mathcal{C}^{(1)}\right]_{rr'}^{-1}\,,
\end{equation}
\begin{equation}
\mathcal{R}_{ss'}^{(1)}(\omega)\equiv
\left[\left(G^{(\infty)}(\omega)\right)^{-1}-\mathcal{D}^{(1)}\right]_{ss'}^{-1}\,,
\end{equation}
and
\begin{equation}
    G^{(\infty)}_{rr'}(\omega) \equiv \frac{\delta_{rr'}}{\omega-E_r^>+i\eta}\,,
\end{equation}
\begin{equation}
    G^{(\infty)}_{ss'}(\omega) \equiv \frac{\delta_{ss'}}{\omega-E_s^<-i\eta}\,.
\end{equation}
\end{subequations}

This notation proves convenient for the derivation of an MCDE system of equations based on the $\mathrm{SRT}(1,1)$ self-energy.

\subsection{Derivation of the (3,1)-MCDE from $\mathrm{SRT}(1,1)$}

To connect the $\mathrm{SRT}(1,1)$ approximation to the Green's function obtained as the solution of the (3,1)-MCDE system of equations, Dyson's equation provides a natural starting point as it relates the self-energy to the Green's function. Replacing the self-energy in Eq.~\eqref{eq:1BDEinf} by its $\mathrm{SRT}(1,1)$ approximation
\begin{equation}
\label{eq:Dyson_recovered}
g_{\alpha\beta}(\omega)=g_{\alpha\beta}^{(\infty)}(\omega)+g_{\alpha\gamma}^{(\infty)}(\omega)\mathcal{M}^{(1)\dagger}_{\gamma r}\,
\mathcal{R}^{(1)}_{rr'}(\omega)\,
\mathcal{M}^{(1)}_{r'\delta}g_{\delta\beta}(\omega)+g_{\alpha\gamma}^{(\infty)}(\omega)\mathcal{N}_{\gamma s}^{(1)}\,
\mathcal{R}_{ss'}^{(1)}(\omega)\,
\mathcal{N}_{s'\delta}^{(1)\dagger}g_{\delta\beta}(\omega)\, ,
\end{equation}
and introducing
\begin{subequations}
 \label{eq:secondmcde}
\begin{equation}
   \begin{aligned}
        {G}^{2e1h/1p}_{r\beta}(\omega) =& \,\mathcal{R}_{rr'}^{(1)}(\omega)\,
\mathcal{M}_{r'\delta}^{(1)}g_{\delta\beta}(\omega)\\
=& \,G^{(\infty)}_{rr'}(\omega)\mathcal{M}_{r'\delta}^{(1)}g_{\delta\beta}(\omega)+G_{rr'}^{(\infty)}(\omega)\mathcal{C}_{r'r''}^{(1)}{G}^{2e1h/1p}_{r''\beta}(\omega)\,.
 \end{aligned}\end{equation}
 \begin{equation}
    \begin{aligned}
        {G}^{1e2h/1p}_{s\beta}(\omega) =& \,\mathcal{R}_{ss'}^{(1)}(\omega)\,
\mathcal{N}_{s'\delta}^{(1)\dagger}g_{\delta\beta}(\omega)\\
=& \,G^{(\infty)}_{ss'}(\omega)\mathcal{N}_{s'\delta}^{(1)\dagger}g_{\delta\beta}(\omega)+G_{ss'}^{(\infty)}(\omega)\mathcal{D}_{s's''}^{(1)}{G}^{1e2h/1p}_{s''\beta}(\omega)\,.
 \end{aligned}\end{equation}
\end{subequations}

Eq.~\eqref{eq:Dyson_recovered} can be written as 
\begin{equation}
    \label{eq:firstmcde}
        g_{\alpha\beta}(\omega)=g_{\alpha\beta}^{(\infty)}(\omega)+g_{\alpha\gamma}^{(\infty)}(\omega)\mathcal{M}_{\gamma r}^{(1)\dagger}\,{G}_{r\beta}^{2e1h/1p}(\omega)+g_{\alpha\gamma}^{(\infty)}(\omega)\mathcal{N}_{\gamma s}^{(1)}\,{G}_{s\beta}^{1e2h/1p}(\omega)\,.
\end{equation}

Equations~\eqref{eq:secondmcde} and~\eqref{eq:firstmcde} constitute the first three building blocks of the (3,1)-MCDE equations. To find the last two, it is convenient to rewrite Eq.~\eqref{eq:Dyson_recovered}  as 
\begin{equation}
\label{eq:greenfunceq}
    g_{\alpha\beta}(\omega) = \left[\left(g_{\alpha\beta}^{(\infty)}(\omega)\right)^{-1}-\mathcal{M}_{\alpha r}^{(1)\dagger}\mathcal{R}^{(1)}_{rr'}(\omega)\mathcal{M}_{r'\beta}^{(1)}-\mathcal{N}_{\alpha s}^{(1)}\mathcal{R}^{(1)}_{ss'}(\omega)\mathcal{N}_{s'\beta}^{(1)\dagger}\right]^{-1}\,.
\end{equation}

By successively expanding the resolvent $\mathcal{R}$ and then the resolvent $g$ in Eq.~\eqref{eq:greenfunceq}, the Green's function can be written as
\begin{equation}
    g_{\alpha\beta}(\omega) = g^{(\infty)}_{\alpha\beta}(\omega)+g_{\alpha \gamma}^{(\infty)}(\omega)\mathcal{M}_{\gamma r}^{(1)\dagger}\mathcal{G}_{rr'}^{2e1h}(\omega)\mathcal{M}_{r'\delta}^{(1)}g_{\delta\beta}^{(\infty)}(\omega)+g_{\alpha \gamma}^{(\infty)}(\omega)\mathcal{N}_{\gamma s}^{(1)}\mathcal{G}_{ss'}^{1e2h}(\omega)\mathcal{N}_{s'\delta}^{(1)\dagger}g_{\delta\beta}^{(\infty)}(\omega)\,,
\end{equation}
with
\begin{subequations}
\begin{equation}
\label{G3P}
    \mathcal{G}_{rr'}^{2e1h}(\omega) =  G_{rr'}^{(\infty)}(\omega) +  G_{rr''}^{(\infty)}(\omega)\Xi_{r''r'''}^{(1)}(\omega)\mathcal{G}_{r'''r'}^{2e1h}(\omega)\,,
\end{equation}
\begin{equation}
\label{G3Pbis}
    \mathcal{G}_{ss'}^{1e2h}(\omega) =  G_{ss'}^{(\infty)}(\omega) +  G_{ss''}^{(\infty)}(\omega)\Xi_{s''s'''}^{(1)}(\omega)\mathcal{G}_{s'''s'}^{1e2h}(\omega)\,,
\end{equation}
where
\begin{equation}
    \Xi^{(1)}_{rr'}(\omega) \equiv \left[\mathcal{C}^{(1)}_{rr'}+\mathcal{M}_{r\alpha}^{(1)}g_{\alpha\beta}^{(\infty)}(\omega)\mathcal{M}_{\beta r'}^{(1)\dagger}\right]\,,
\end{equation}
\begin{equation}
    \Xi_{ss'}^{(1)}(\omega) \equiv \left[\mathcal{D}_{ss'}^{(1)}+\mathcal{N}^{(1)\dagger}_{s\alpha}g_{\alpha\beta}^{(\infty)}(\omega)\mathcal{N}^{(1)}_{\beta s'}\right]\,.
\end{equation}
\end{subequations}

Introducing 
\begin{subequations}
\label{eq:thirdmcde}
\begin{equation}
    {G}^{1p/2e1h}_{\alpha r}(\omega) \equiv g_{\alpha\beta}^{(\infty)}(\omega)\mathcal{M}_{\beta r'}^{(1)\dagger}\mathcal{G}_{r'r}^{2e1h}(\omega)\,,
\end{equation}
\begin{equation}
    {G}^{1p/1e2h}_{\alpha s}(\omega) \equiv g_{\alpha\beta}^{(\infty)}(\omega)\mathcal{N}^{(1)}_{\beta s'}\mathcal{G}^{1e2h}_{s' s}(\omega)\,,
\end{equation}
\end{subequations}
Eq.~\eqref{G3P} can be written as
\begin{subequations}
\label{eq:fourthmcde}
\begin{equation}\begin{aligned}
    \mathcal{G}_{rr'}^{2e1h}(\omega) =&  \,G_{rr'}^{(\infty)}(\omega) +  G_{rr''}^{(\infty)}(\omega)\mathcal{C}_{r''r'''}^{(1)}\,\mathcal{G}_{r'''r'}^{2e1h}(\omega)\\
    &+G_{rr''}^{(\infty)}(\omega)\mathcal{M}^{(1)}_{r''\alpha}{G}^{1p/2e1h}_{\alpha r'}(\omega)\,,
\end{aligned}\end{equation}
\begin{equation}\begin{aligned}
    \mathcal{G}_{ss'}^{1e2h}(\omega) =&  \,G_{ss'}^{(\infty)}(\omega) +  G_{ss''}^{(\infty)}(\omega)\mathcal{D}_{s''s'''}^{(1)}\,\mathcal{G}_{s'''s'}^{1e2h}(\omega)\\
    &+G_{ss''}^{(\infty)}(\omega)\mathcal{N}^{(1)\dagger}_{s''\alpha}{G}^{1p/1e2h}_{\alpha s'}(\omega)\,,
\end{aligned}\end{equation}
\end{subequations}
Equations~\eqref{eq:thirdmcde} and~\eqref{eq:fourthmcde} complete the set of (3,1)-MCDE equations. Together with Eqs.~\eqref{eq:secondmcde} and~\eqref{eq:firstmcde}, they can be recast into a single (3,1)-MCDE matrix equation
\begin{comment}   
\begin{equation}
\label{eq:basicMCDEequation2}
\begin{aligned}
\begin{pmatrix}
g_{\alpha\beta} & G^{1p/2e1h}_{\alpha r'} & G^{1p/1e2h}_{\alpha s'}\\
\tilde{G}^{2e1h/1p}_{r\beta} & \mathcal{G}^{2e1h}_{rr'} & 0\\
\tilde{G}^{1e2h/1p}_{s\beta} & 0 & \mathcal{G}^{1e2h}_{ss'}
\end{pmatrix}
=&
\begin{pmatrix}
g^{\mathrm{(\infty)}}_{\alpha\beta} & 0 & 0 \\
0 & G^{(\infty)}_{rr'} & 0\\
0 & 0 & G^{(\infty)}_{ss'}
\end{pmatrix}
\\
&+
\begin{pmatrix}
g^{\mathrm{(\infty)}}_{\alpha\gamma} & 0 & 0 \\
0 & G^{(\infty)}_{rr''} & 0\\
0 & 0 & G^{(\infty)}_{ss''}
\end{pmatrix}
\begin{pmatrix}
0 & \mathcal{M}^{(1)\dagger}_{\gamma r'''} & \mathcal{N}^{(1)}_{\gamma s'''} \\
\mathcal{M}^{(1)}_{r''\delta} & \mathcal{C}^{(1)}_{r''r'''} & 0 \\
\mathcal{N}^{(1)\dagger}_{s''\delta} & 0 & \mathcal{D}^{(1)}_{s''s'''}
\end{pmatrix}
\begin{pmatrix}
g_{\delta\beta} & G^{1p/2e1h}_{\delta r'} & G^{1p/1e2h}_{\delta s'}\\
\tilde{G}^{2e1h/1p}_{r'''\beta} & \mathcal{G}^{2e1h}_{r'''r'} & 0\\
\tilde{G}^{1e2h/1p}_{s'''\beta} & 0 & \mathcal{G}^{1e2h}_{s'''s'}
\end{pmatrix}\, ,
\end{aligned}
\end{equation}
\end{comment}
\begin{subequations}
\label{eq:basicMCDEequation2}
\setlength{\arraycolsep}{3pt}
\renewcommand{\arraystretch}{1.2}
\begin{align}
  \mathbf{G} &= \mathbf{G}^{(\infty)} + \mathbf{G}^{(\infty)}\, \boldsymbol{\Sigma}\, \mathbf{G} \, ,
  \label{eq:basicMCDEequation2} \\
  \intertext{with the block matrices}
  \mathbf{G} &=
  \begin{pmatrix}
g_{\alpha\beta} & G^{1p/2e1h}_{\alpha r'} & G^{1p/1e2h}_{\alpha s'}\\
{G}^{2e1h/1p}_{r\beta} & \mathcal{G}^{2e1h}_{rr'} & 0\\
{G}^{1e2h/1p}_{s\beta} & 0 & \mathcal{G}^{1e2h}_{ss'}
\end{pmatrix} \,,
  \label{eq:MCDEblockG} \\[2ex]
  \mathbf{G}^{(\infty)} &=
  \begin{pmatrix}
g^{\mathrm{(\infty)}}_{\alpha\beta} & 0 & 0 \\
0 & G^{(\infty)}_{rr'} & 0\\
0 & 0 & G^{(\infty)}_{ss'}
\end{pmatrix} \,,
  \label{eq:MCDEblockG0} \\[2ex]
   \boldsymbol{\Sigma} &=
  \begin{pmatrix}
0 & \mathcal{M}^{(1)\dagger}_{\gamma r'''} & \mathcal{N}^{(1)}_{\gamma s'''} \\
\mathcal{M}^{(1)}_{r''\delta} & \mathcal{C}^{(1)}_{r''r'''} & 0 \\
\mathcal{N}^{(1)\dagger}_{s''\delta} & 0 & \mathcal{D}^{(1)}_{s''s'''}
\end{pmatrix}\,.
  \label{eq:MCDEblockSig}
\end{align}
\end{subequations}

Starting from Dyson's equation~Eq.~\eqref{eq:Dyson_recovered} in the $\mathrm{SRT}(1,1)$ approximation, a Baranger based (3,1)-MCDE set of equations is recovered. Thus, the one-body Green's function solution of Eq.~\eqref{eq:Dyson_recovered} also satisfies the (3,1)-MCDE set of equations. As a result, the (3,1)-MCDE approximation lies between the ADC(2) and ADC(3) truncation order.

\section{R-HF and U-HF reference state in the Hubbard model}
\label{app:symr-symb}

The use of the ADC and MCDE approximations presupposes the existence of a nondegenerate single-reference HF state with respect to particle-hole excitations. For lattices with $L = 4M$, $M = 1, \cdots $, R-HF preserves the degeneracy of the frontier orbitals at the Fermi level, resulting in multiple energetically equivalent Slater determinants. In contrast, U-HF lifts this degeneracy through spin symmetry breaking, leading to a unique lowest-energy broken-symmetry Slater determinant. 

\subsection{Symmetries of the exact non-magnetic ground state}

The one-dimensional Hubbard model with repulsive on-site interaction ($U > 0$), an even number $L$ of lattice sites and periodic boundary conditions is considered at
half-filling. Under these conditions, Lieb's theorem~\cite{Lieb1989,Tasaki1998,Tasaki2020} applies 
to guarantee that the ground state of the Hubbard Hamiltonian is unique 
(nondegenerate) and has total spin $S = 0$, i.e.\ it is an SU(2) 
singlet. As a consequence of the SU(2) spin-rotation invariance of the 
Hamiltonian and the singlet nature of the ground state, no net local spin 
polarization can occur. The ground state is said to be non-magnetic (NM),
which implies that the spin-resolved local densities satisfy
\[
\langle n_{i\uparrow} \rangle = \langle n_{i\downarrow} \rangle
\qquad \forall i \in \{1,2,\ldots,L\}.
\]
In addition, the Hamiltonian is translationally invariant due to the homogeneity of the lattice and the periodic boundary conditions. Since the ground state is unique, this invariance enforces spatial uniformity of all local observables. In particular, the local charge density must be site independent~\cite{Essler2005},
\[
\langle n_i \rangle = \langle n_j \rangle
\qquad \forall i,j\in \{1,2,\ldots,L\}.
\]
At half-filling, the Hubbard Hamiltonian is furthermore invariant under a particle-hole transformation. This symmetry fixes the average occupation per site~\cite{Essler2005,Tasaki1998,Gebhard1997} and imposes
\[
\langle n_i \rangle = \langle n_{i\uparrow} \rangle + \langle n_{i\downarrow} \rangle = 1
\qquad \forall i\in \{1,2,\ldots,L\}.
\]
Combining translational invariance, particle-hole symmetry, and $SU(2)$ spin symmetry, one finally obtains a uniform spin-resolved density in the exact ground state,
\begin{equation}
\left\langle n_{i\sigma} \right\rangle = \frac{1}{2}
\qquad
\forall i \in \{1,2,\ldots,L\}, \ \forall \sigma \in \{\uparrow,\downarrow\}.
\end{equation}

\subsection{Open-shell character of R-HF}
\label{sec:open}

The HF Slater determinant is built by occupying $L$ eigenstates of the HF one-body Hamiltonian~\cite{Hubbard1963,Fazekas1999}
\begin{equation}
\label{eq:HF}
\begin{aligned}
    h^{(\mathrm{HF})} 
    =& -t\sum_{i=1}^{L}\sum_{\sigma=\uparrow,\downarrow} 
       \left(c_{i\sigma}^\dagger c_{(i+1)\sigma} + c_{(i+1)\sigma}^\dagger c_{i\sigma}\right) \\
    & + U\sum_{i=1}^L \left(\langle n_{i\uparrow}\rangle n_{i\downarrow} + n_{i\uparrow}\langle n_{i\downarrow}\rangle\right) 
    + \varepsilon_0 \sum_{i=1}^L \sum_{\sigma=\uparrow,\downarrow} n_{i\sigma},
\end{aligned}
\end{equation}
where $\langle n_{i\sigma}\rangle \equiv \langle \Phi^{(\mathrm{HF})}| 
c^\dagger_{i\sigma} c_{i\sigma} |\Phi^{(\mathrm{HF})}\rangle$ is the local density 
at site $i$ for spin $\sigma$, evaluated self-consistently in the HF 
Slater determinant $|\Phi^{(\mathrm{HF})}\rangle$. 

Motivated by the symmetries of the exact ground state, we construct a 
symmetry-restricted HF (R-HF) reference state, assuming
\begin{equation}
    \langle n_{i\sigma}\rangle = \frac{1}{2} \quad \forall\, (i,\sigma).
\end{equation}
The HF Hamiltonian then simplifies to
\begin{equation}
\begin{aligned}
    h^{(\mathrm{HF})} =& -t\sum_{i=1}^{L}\sum_{\sigma=\uparrow,\downarrow} 
    \left(c_{i\sigma}^\dagger c_{(i+1)\sigma} + \text{h.c.}\right)
    \\
    &+ \left(\frac{U}{2} + \varepsilon_0\right)\sum_{i=1}^L \sum_{\sigma=\uparrow,\downarrow} n_{i\sigma}.
\end{aligned}
\end{equation}
By choosing $\varepsilon_0 = -U/2$, the on-site term is exactly canceled at half-filling, so that the HF Hamiltonian reduces to its kinetic contribution. This choice ensures that the HF Hamiltonian is explicitly particle–hole symmetric.

The HF single-particle eigenvalues for a ring of $L$ sites are then
\begin{equation}
    \varepsilon_n = -2t \cos\!\left(\frac{2\pi n}{L}\right), \qquad n = 0, \ldots, L-1.
\end{equation}
If $L = 4M$, with $M \in \mathbb{N}$, there exist two indices
\begin{equation}
    m_1 = M, \qquad m_2 = 3M,
\end{equation}
such that
\begin{equation}
    \varepsilon_{m_1} = \varepsilon_{m_2} = 0.
\end{equation}

Thus, when $L = 4M$, the half-filled R-HF reference is degenerate: two single-particle levels are degenerate at zero energy, but only one of them can be occupied at half-filling. The $4M$-sites Hubbard ring is then said to be open-shell. The ADC(2) and (3,1)-MCDE approximations rely on a nondegenerate reference state to generate well-defined excitations. In this degenerate case, the occupation of the frontier levels is not unique, so the R-HF ground state is not a single well-defined determinant, leading to ambiguous particle-hole states. Consequently, the construction of the ADC(2) or (3,1)-MCDE matrices becomes ill-defined: the Dyson matrices and the associated coupled equations cannot be properly formulated. Therefore, the degeneracy of the R-HF reference at half-filling for $L = 4M$ prevents the straightforward application of the ADC(2) /(3,1)-MCDE approximations. This obstruction is lifted by a symmetry-broken U-HF reference, which removes the degeneracy and provides a single nondegenerate determinant on which the spectral function is computed.
%In the next section, we will allow the reference state to break the spin symmetry. 
\end{widetext}

\end{document}